\documentclass[aps,prd,a4paper,onecolumn,amsmath,showpacs,superscriptaddress,nofootinbib,preprintnumbers,notitlepage]{revtex4-1}
\usepackage{amssymb,amsmath}
\usepackage{graphicx}
\usepackage{color}
\usepackage{dcolumn}
\usepackage{pgfplots}
\usepackage{booktabs}
\usepackage{multirow}
\usepackage{dcolumn}
\usepackage{amsmath}
\usepackage{mathtools}
\usepackage{amsfonts}
\usepackage{amssymb}
\usepackage{epstopdf}
\usepackage{bm}
\usepackage{siunitx}
\usepackage{braket}
\usepackage{enumitem}
\usepackage{soul}
\usepackage{ulem}
\usepackage{color}
\usepackage{transparent}
\usepackage{pifont}
\usepackage[font={small}]{caption}
\newcommand\rf[1]{(\ref{eq:#1})}
\newcommand\lab[1]{\label{eq:#1}}
\newcommand\nonu{\nonumber}
\newcommand\br{\begin{eqnarray}}
\newcommand\er{\end{eqnarray}}
\newcommand\be{\begin{equation}}
\newcommand\ee{\end{equation}}

\newcommand\lb{\lbrack}
\newcommand\rb{\rbrack}

\renewcommand\({\left(}
\renewcommand\){\right)}

\newcommand\bc{\begin{center}}
\newcommand\ec{\end{center}}

\newcommand\partder[2]{\frac{{\partial {#1}}}{{\partial {#2}}}}

\renewcommand\a{\alpha}

\renewcommand\d{\delta}

\newcommand\eps{\epsilon}
\newcommand\vareps{\varepsilon}

\newcommand\G{\Gamma}

\newcommand\h{\frac{1}{2}}
\renewcommand\k{\kappa}
\renewcommand\l{\lambda}

\newcommand\m{\mu}
\newcommand\n{\nu}
\newcommand\om{\omega}
\renewcommand\O{\Omega}

\newcommand\vp{\varphi}
\renewcommand\P{\Phi}
\newcommand\pa{\partial}

\renewcommand\th{\theta}

\newcommand\cF{{\mathcal F}}

\newcommand\cP{{\mathcal P}}

\newcommand{\ct}[1]{\cite{#1}}
\newcommand{\bib}[1]{\bibitem{#1}}

\newcommand\PRL[3]{\textsl{Phys. Rev. Lett.} \textbf{#1} (#2) #3}

\newcommand\PRD[3]{\textsl{Phys. Rev.} \textbf{D#1} (#2) #3}

\newcommand\PLB[3]{\textsl{Phys. Lett.} \textbf{#1B} (#2) #3}
\newcommand\CQG[3]{\textsl{Class. Quantum Grav.} \textbf{#1} (#2) #3}

\newcommand\PR[3]{\textsl{Phys. Reports} \textbf{#1} (#2) #3}

\newcommand\IJMPD[3]{\textsl{Int. J. Mod. Phys.} \textbf{D#1} (#2) #3}

\newcommand\MPLA[3]{\textsl{Mod. Phys. Lett.} \textbf{A#1} (#2) #3}

\newcommand\vpdot{\stackrel{.}{\varphi}}
\newcommand\vpddot{\stackrel{..}{\varphi}}

\newcommand\adot{\stackrel{.}{a}}
\newcommand\addot{\stackrel{..}{a}}
\newcommand\Hdot{\stackrel{.}{H}}

\begin{document}
\title { Unification: Emergent universe followed by inflation and dark epochs from multi-field  theory.
}

\author{Eduardo Guendelman}
\email{guendel@bgu.ac.il}
\affiliation{Department of Physics, Ben-Gurion University of the Negev, Beer-Sheva, Israel.\\}
\affiliation{Frankfurt Institute for Advanced Studies (FIAS),
Ruth-Moufang-Strasse 1, 60438 Frankfurt am Main, Germany.\\}
\affiliation{Bahamas Advanced Study Institute and Conferences, 
4A Ocean Heights, Hill View Circle, Stella Maris, Long Island, The Bahamas.
}

\author{Ram\'{o}n Herrera}
\email{ramon.herrera@pucv.cl}
\affiliation{Instituto de F\'{\i}sica, Pontificia Universidad Cat\'{o}lica de Valpara\'{\i}so, Avenida Brasil 2950, Casilla 4059, Valpara\'{\i}so, Chile.
}
\begin{abstract}
A two scalar field model that incorporates non Riemannian Measures of integration  or usually called Two Measures Theory (TMT)  is introduced, in order to unify the early and present universe. In the Einstein frame a K-essence is generated and as a consequence for the early universe, we can have a  Non Singular Emergent universe followed by Inflation and for the present universe dark epochs with consistent generation of dark energy (DE), dark matter (DM) and stiff matter. The  scale invariance is introduced and then is spontaneously broken from the integration of the degrees of freedom associated with the modified measures. The resulting effective potentials and K-essence  in the Einstein frame produce three flat regions corresponding to the different epochs mentioned before. For the first flat region we can associate  an emergent and an inflationary universe. Here for a parameter-space region this flat plateau  possesses a non singular stable emergent universe solution which characterizes   an initial  epoch of evolution that precedes the inflationary scenario. Also assuming this first plateau, we study the inflation in the framework of the slow-roll approximation. In this scenario under the slow roll approximation we obtain a linear combination that is a constant. The corresponding cosmological perturbations in our model are determined and we also obtain the different constrains on the parameter-space  from the Planck data.

In the following  flat region DE and also the DM, which does not need to be introduced separately, it is instead a result of a K-essence induced by the multi measures, multi field theory. Also stiff matter component is automatically generated from the K-essence theory from two scalar fields. From the perturbative analysis associated to the perturbation  solution of background, we find a correlation  between the two scalars. Besides, we obtain that our model during the  
dark epoch has a behaviour of tracking freezing model.

\end{abstract}
\maketitle
\section{Introduction}
\label{intro}

In the most popular paradigm  for the early universe, it is postulated that 
the universe suffers a period of exponential expansion called ``inflation'' introduced by Guth \cite{Guth} , Starobinsky \cite {starobinsky} and others (cf. the books \ct{early-univ,primordial} and references therein). However   these models do not address  the big-bang singularity problem. Nevertheless  a scenario that while incorporating inflation, takes into account of an even earlier period, called  the emergent period or emergent universe was proposed in Refs.\cite{emerg,emerg2}.  
On the other hand, the late universe after the discovery of the accelerating universe  \ct{accel-exp,accel-exp-2}, also shows a period that some common features with the inflationary period, although the relevant scales are very different. In this framework,  the late universe ``standard cosmological'' description  for the late universe, is now the   $\Lambda$CDM  model \ct{lambdaCDM}. This model incorporates  a cosmological constant and a  Dark Matter (DM) consisting of pressure-less dust and ordinary visible matter (which is also dust) and in this scenario  
the universe being now dominated by the Cosmological Constant  $\Lambda$ or Dark Energy (DE).
The $\Lambda$CDM model is now  being challenged by the discovery of some observational tensions, the $H_0$  and the $\sigma_8$ tensions \ct{H0,sigma8}.

In the inflationary period also primordial density perturbations are
generated (Ref.\ct{primordial} and references therein). The ``inflation'' 
is followed by particle creation, where the observed  matter and radiation were generated
\ct{early-univ}, and finally the evolution arrives to a present phase of slowly 
accelerating universe \ct{accel-exp,accel-exp-2}.
In this standard model, however, at least three fundamental questions remain 
unanswered:  

\begin{itemize}
\item 
 The  inflationary epoch, although solving many cosmological puzzles, like the
 horizon,  flatness problems etc. and also providing a mechanism for primordial density perturbations, cannot address the initial singularity 
problem; 
\item 
In this model there is no reason  for the existence of two scenarios  of 
exponential expansion with such wildly different scales the inflationary 
epoch and the current era of slowly accelerated expansion of the universe.

\item 
There is no explanation for the existence of the Dark Matter, that represents an invisible part and most of the dust in the universe that must be different from the baryonic matter, the visible part of the dust component of our universe.
\end{itemize}
The best known mechanism for producing a phase  of accelerated expansion 
is from   some vacuum energy or a cosmological constant. In the framework of a standard   
scalar field theory, vacuum energy density arises  when the scalar
field acquires an effective potential $U_{\rm eff}$ which has flat plateaus  so 
that the  field (inflaton) can ``slowly roll'' \ct{slow-roll,slow-roll-param} and its 
kinetic energy can be disregarded  with which  the energy-momentum tensor 
$T_{\m\n} \simeq - g_{\m\n} U_{\rm eff}$, a contribution that mimics that of a cosmological constant.
We will make use of this mechanism to explain the accelerated expansion of the universe associated to the inflation phase of the universe. In the late phases of the universe we will also find a contribution from an effective scalar field potential, 
but there will be also a contribution to the DE from a K-essence background configuration that also gives rise to a contribution that satisfies the equation of state $p=-\rho$.
Besides, in the emergent universe the effective potential contribution also can combine with the K-essence to produce  
 an equation of state $ p= -\rho/3$ which produces no acceleration and in particular a static Einstein universe, which is stable for a range of parameters, unlike the original Einstein universe. Therefore the interplay of an effective potential and K-essence will be crucial for many effects studied in this paper along many continuously connected phases of the universe.

The connection  between the
inflationary epoch to a slowly 
accelerating universe through the evolution of a single scalar field denominated  ``the
 quintessential inflation scenario''
 has been first analyzed  in 
Ref.\ct{peebles-vilenkin}. 
Besides,  a quintessential inflation mechanism developed  on 
the K-essence  model, was studied in  Ref.\ct{saitou-nojiri}. Also   quintessential inflation  founded  on the 
``variable gravity'' model was  developed in Ref.\ct{wetterich} and for another  list of references, 
see Ref.\ct{murzakulov-etal}. Additionally other approaches  founded  on the so called $\alpha$ attractors, which utilizes  non canonical kinetic terms have been analyzed  in Ref.\ct{alphaatractors}. Also a quintessential inflation developed  on a  Lorentzian slow- roll approximation  which automatically gives two flat regions was studied in \ct{Lorentzian}.
In addition, 
the $F(R)$ models can connect 
both periods  an early time inflationary era and a late time de Sitter epoch from different values of effective vacuum energies, see  Ref.\ct{starobinsky-2}.
For a review of $f(R)$ model and another modified gravity, see Refs.\cite{Re1, Re2}.

The framework for this research is the use of the metric independent non Riemannian measures for the construction of modified gravity theories
Refs.\ct{TMT-orig-1}-\ct{TMT-orig-3} (see also 
Refs.\ct{TMT-recent-1-a}-\ct{TMT-recent-2}), in some instances we have included the standard measure as well, where the standard Riemannian integration measure  might also contain a Weyl-scale symmetry preserving $R^2$-term \ct{TMT-orig-3}. Some applications have been, (i) $D=4$-dimensional models of gravity and matter fields containing  the new measure of integration appear to be promising candidates for resolution  of the dark energy and dark matter problems, the fifth force problem, and a natural mechanism for spontaneous breakdown of global Weyl-scale symmetry \ct{TMT-orig-1}-\ct{TMT-recent-2}, (ii) To study of reparametrization invariant theories of extended objects (strings and branes) based on employing of a modified non-Riemannian 
world-sheet/world-volume integration measure \ct{mstring}, \ct{nishino-rajpoot}, leads to dynamically 
induced variable string/brane tension and to string models of non-abelian 
confinement, interesting consequences from the modified measures spectrum \ct{mstringspectrum}, and construction of new braneworld scenarios \ct{mstringbranes}, (iii) To study in Ref.\ct{susy-break} of modified supergravity models with an alternative non-Riemannian volume form on the space-time manifold .

Directly connected to the present research are our  papers \ct{ourquintessence} where we have studied  
a unified scenario where both an inflation 
and a slowly accelerated phase for the universe can appear naturally from the 
existence of two flat regions in the effective scalar field potential which
we derive systematically from a Lagrangian action principle. 
Namely, we started with a new kind of globally Weyl-scale invariant gravity-matter 
action within the first-order (Palatini) approach formulated in terms of two 
different non-Riemannian volume forms (integration measures) \ct{quintess}. 
Also we have studied the reheating scenario in this theory assuming the curvaton field in order to decay to the hot big bang model\cite{curv,curv2}.
In this new theory there is a single scalar field with kinetic terms coupled to 
both non-Riemannian measures, and in addition to the scalar curvature 
term $R$ also an $R^2$ term is included (which is similarly allowed by global 
Weyl-scale invariance). Scale invariance is spontaneously broken upon solving part 
of the corresponding equations of motion due to the appearance of two 
arbitrary dimensionfull integration constants. Furthermore, in a subsequent paper we generalized this to a two-field case\ct{ourquintessencewithEDE}
where three flat regions appear, one for inflation and the remaining two for early DE and a late DE phase. In order to describe the early DE and DM, we have introduced a matter action defined as a scale invariant form, which is independent  of the scalar fields.

In this paper we will add a new aspect, that is, without introducing any new additional matter action, therefore not introducing the DM  separately,   as compared to \ct{ourquintessencewithEDE}, we consider instead more generic scale invariant couplings, like involving $R^2$ and others, and we produce as a result a theory  that is also able to provide a non singular emergent phase for the universe where the effective potential contribution  combines with the K-essence to produce a 
$ p= -\rho/3$ equation of state which produces no acceleration and in particular a static Einstein Universe, which is stable for a range of parameters, unlike the original Einstein universe. This emergent phase is then followed by the inflationary phase as well as,   in addition after the emergent and inflation eras we have a  second flat region   that provide 
an explanation for the DE and DM of the late universe.  This is a consequence of the K-essence produced by  the additional kinetic  terms and $ R$ square terms introduced into the action. In this context, we have a contribution to the DE that comes from an effective scalar field potential and also there is a contribution to the DE from a K-essence background configuration. Aditionally we have a  DM component   and an additional stiff matter component which originate from the perturbation of the background K-essence configuration. Here we obtain a consistency condition   that correlates the perturbations of the two fields in the late universe. Also, the different constraints from each phase are shown to be consistent with the constraints on the same parameters from the other phases.

 The plan of the present article is as follows. In the next Section  we discussion  
the general formalism for the new class of gravity-matter considering  two independent non Riemannian volume forms. 
In Section \ref{flat-regions} we analyze the 
three flat regions from the effective scalar potential
in the 
 Einstein framework. 
In Section \ref{cuatro} we study the non singular emergent universe. Also we determine the stability of the solution in a region of the parameter-space.

In Section \ref{inflacion} we describe the inflationary epoch. Here we analyze the slow-roll approximation during the background scenario and also we determine the perturbations cosmological. In this context we find the constraints of the different parameters from the Planck data.
 In Section \ref{DEDM}, we analyze how our model of K-essence can describe the late universe. From the second flat plateau we find different components assuming the perturbations K-essence background configuration. Also we determine the equation of state associated to the dark sector together with the derivative of this parameter. 
 In Section \ref{discuss}  
 we end up with a discussion section concerning generalizations of the model, additional mechanisms for inducing DE and the role of the third flat region and its possible relation to early DE.
 For simplicity 
 we will consider units where the Newton constant is taken as $G_{\rm Newton} = 1/16\pi$.

\section{Theory using Two Independent Non-Riemannian Volume-Forms}
\label{TMMT}

We shall assume the  action 
of the general form involving two independent non-metric integration
measure densities generalizing the model analyzed in \ct{quintess} is given by 
\be
S = \int d^4 x\,\P_1 (A) \Bigl\lb R + L^{(1)} \Bigr\rb +  
\int d^4 x\,\P_2 (B) \Bigl\lb L^{(2)} + \eps R^2 + 
\frac{\P (H)}{\sqrt{-g}}\Bigr\rb \; .
\lab{TMMT1}
\ee
Here the following definitions  are used:

\begin{itemize}
\item
The quantities $\P_{1}(A)$ and $\P_2 (B)$ are two densities and these are  independent non-metric volume-forms defined in terms of field-strengths of two auxiliary 3-index antisymmetric
tensor gauge fields
\be
\P_1 (A) = \frac{1}{3!}\vareps^{\m\n\k\l} \pa_\m A_{\n\k\l} \quad ,\quad
\P_2 (B) = \frac{1}{3!}\vareps^{\m\n\k\l} \pa_\m B_{\n\k\l} \; .
\lab{Phi-1-2}
\ee

\item
The scalar curvature $R = g^{\m\n} R_{\m\n}(\G)$ and the Ricci tensor $R_{\m\n}(\G)$ are defined in the first-order (Palatini) formalism, in which the affine
connection $\G^\m_{\n\l}$ is \textsl{a priori} independent of the metric $g_{\m\n}$.
 Let us recall that $R+R^2$ gravity within the
second order formalism  was originally
developed in \ct{starobinsky}.
\item
The two different Lagrangians $L^{(1,2)}$ correspond to two  scalar matter fields $\varphi_1$ and $\varphi_2$ such that
\br
L^{(1)} = -\h g^{\m\n} \pa_\m \vp_1 \pa_\n \vp_1 -\h g^{\m\n} \pa_\m \vp_2 \pa_\n \vp_2- V(\vp_1,\vp_2),\quad
\; 
\lab{L-11} \\
L^{(2)} = -\frac{b_1}{2} e^{-\a_1\vp_1} g^{\m\n} \pa_\m \vp_1 \pa_\n \vp_1 -\frac{b_2}{2} e^{-\a_2\vp_2} g^{\m\n} \pa_\m \vp_2 \pa_\n \vp_2+  U(\vp_1,\vp_2) 
 ,\quad 
\lab{L-2}
\er
where the potentials $V(\vp_1,\vp_2)$ and $U(\vp_1,\vp_2)$ are defined as 
\be
V(\vp_1,\vp_2) = f_1 \exp \{-\a_1\vp_1\} +g_1 \exp \{-\a_2\vp_2\},\,\,\mbox{and}\,\,\,\,U(\vp_1,\vp_2) = f_2 \exp \{-2\a_1\vp_1\}+g_2 \exp \{-2\a_2\vp_2\} .
\lab{U,V}
\ee
Here $\a_1,\a_2, f_1,g_1, f_2, \epsilon$ and $g_2$ are  positive parameters, whereas $b_1$ and $b_2$ are
dimensionless and their  signs are to be discussed. The parameters $\alpha_1$ and $\alpha_2$ have  dimensions of $M_{Pl}^{-1}$, instead  the parameters $f_1$,  $f_2$, $g_1$ and $g_2$ have units of $M_{Pl}^4$ and  the parameter $\epsilon$ has units of $M_{Pl}^{-2}$. Let us recall that since we are considering units in which $G_{\rm Newton} = 1/16\pi$ then the Planck mass $M_{Pl}=\sqrt{2}=\sqrt{1/8\pi G_{\rm Newton}}$.

\item
The density $\P (H)$ denotes  the dual field strength of a third auxiliary 3-index antisymmetric
tensor 
\be
\P (H) = \frac{1}{3!}\vareps^{\m\n\k\l} \pa_\m H_{\n\k\l} \; .
\lab{Phi-H}
\ee
\end{itemize}

The action given by Eq.\rf{TMMT1} is invariant under the global Weyl-scale transformations: 
\br
g_{\m\n} \to \l g_{\m\n} \;\; ,\;\; \G^\m_{\n\l} \to \G^\m_{\n\l} \;\; ,\;\; 
\vp_1 \to \vp_1 + \frac{1}{\a_1}\ln \l \;\;,\vp_2 \to \vp_2 + \frac{1}{\a_2}\ln \l 
\nonu \\
A_{\m\n\k} \to \l A_{\m\n\k} \;\; ,\;\; B_{\m\n\k} \to \l^2 B_{\m\n\k}
\;\; ,\;\; H_{\m\n\k} \to H_{\m\n\k} \; .
\lab{scale-transf1}
\er

The variation of the action \rf{TMMT1} w.r.t. affine connection $\G^\m_{\n\l}$ we have
\be
\int d^4\,x\,\sqrt{-g} g^{\m\n} \Bigl(\frac{\P_1}{\sqrt{-g}} +
2\eps\,\frac{\P_2}{\sqrt{-g}}\, R\Bigr) \(\nabla_\k \d\G^\k_{\m\n}
- \nabla_\m \d\G^\k_{\k\n}\) = 0, 
\lab{var-G}
\ee
and following Ref.\ct{TMT-orig-1}, its solution 
$\G^\m_{\n\l}$ becomes a Levi-Civita connection 
\be
\G^\m_{\n\l} = \G^\m_{\n\l}({\bar g}) = 
\h {\bar g}^{\m\k}\(\pa_\n {\bar g}_{\l\k} + \pa_\l {\bar g}_{\n\k} 
- \pa_\k {\bar g}_{\n\l}\) \; ,
\lab{G-eq}
\ee
where the Weyl-rescaled metric ${\bar g}_{\m\n}$ is given by
\be
{\bar g}_{\m\n} = (\chi_1 + 2\eps \chi_2 R) g_{\m\n} \;\; ,\;\;\mbox{where}\,\,\,\, 
\chi_1 \equiv \frac{\P_1 (A)}{\sqrt{-g}} \;\; ,\;\;\mbox{and}\,\,\,\,
\chi_2 \equiv \frac{\P_2 (B)}{\sqrt{-g}} \; .
\lab{bar-g1}
\ee

On the other hand, the variation of  \rf{TMMT1} w.r.t. auxiliary tensors 
$A_{\m\n\l}$, $B_{\m\n\l}$ and $H_{\m\n\l}$ becomes
\be
\pa_\m \Bigl\lb R + L^{(1)} \Bigr\rb = 0 \quad, \quad
\pa_\m \Bigl\lb L^{(2)} + \eps R^2 + \frac{\P (H)}{\sqrt{-g}}\Bigr\rb = 0 
\quad, \quad \pa_\m \Bigl(\frac{\P_2 (B)}{\sqrt{-g}}\Bigr) = 0 \; ,
\lab{A-B-H-eqs}
\ee
whose solutions are
\be
\frac{\P_2 (B)}{\sqrt{-g}} \equiv \chi_2 = {\rm const} \;\; ,\;\;
R + L^{(1)} = - M_1 = {\rm const} \;\; ,\;\; 
L^{(2)} + \eps R^2 + \frac{\P (H)}{\sqrt{-g}} = - M_2  = {\rm const} \; .
\lab{integr-const1}
\ee
Here the parameters $M_1$ and $M_2$ are arbitrary dimensionful and the quantity $\chi_2$ corresponds to an
arbitrary dimensionless integration constant. 
Note that the 
constant $\chi_2$ in \rf{integr-const1} preserves
global Weyl-scale invariance \rf{scale-transf1}, whereas 
the appearance of the second and third integration constants $M_1,\, M_2$
implies 
breakdown of global Weyl-scale invariance 
under \rf{scale-transf1}.

Now the variation of the action  \rf{TMMT1} w.r.t. $g_{\m\n}$ and considering  relations \rf{integr-const1} 
yields 
\be
\chi_1 \Bigl\lb R_{\m\n} + \h\( g_{\m\n}L^{(1)} - T^{(1)}_{\m\n}\)\Bigr\rb -
\h \chi_2 \Bigl\lb T^{(2)}_{\m\n} + g_{\m\n} \(\eps R^2 + M_2\)
- 2 R\,R_{\m\n}\Bigr\rb = 0 \; ,
\lab{pre-einstein-eqs1}
\ee
in which  $T^{(1,2)}_{\m\n}(\vp_1,\vp_2)=T^{(1,2)}_{\m\n}$ are the energy-momentum tensors of the scalar
fields Lagrangians with the standard definitions:
\be
T^{(1,2)}_{\m\n} = g_{\m\n} L^{(1,2)} - 2 \partder{}{g^{\m\n}} L^{(1,2)} \; .
\lab{EM-tensor1}
\ee

By considering  the trace of Eq.\rf{pre-einstein-eqs1} and using again second relation 
\rf{integr-const1} we find  for the scale factor $\chi_1$ the expression 
\be
\chi_1 = 2 \chi_2 \frac{T^{(2)}/4 + M_2}{L^{(1)} - T^{(1)}/2 - M_1} \; ,
\lab{chi-11}
\ee
where the trace $T^{(1,2)} = g^{\m\n} T^{(1,2)}_{\m\n}$. 

From Eqs.\rf{integr-const1} and \rf{pre-einstein-eqs1} we obtain 
 the Einstein-like form:
\br
R_{\m\n} - \h g_{\m\n}R = \h g_{\m\n}\(L^{(1)} + M_1\)
+ \frac{1}{2\O}\(T^{(1)}_{\m\n} - g_{\m\n}L^{(1)}\)
+ \frac{\chi_2}{2\chi_1 \O} \Bigl\lb T^{(2)}_{\m\n} + 
g_{\m\n} \(M_2 + \eps(L^{(1)} + M_1)^2\)\Bigr\rb \; ,
\lab{einstein-like-eqs1}
\er
in which the function $\Omega$ is defined as
\be
\O = 1 - \frac{\chi_2}{\chi_1}\,2\eps\(L^{(1)} + M_1\) \; .
\lab{Omega-eqN}
\ee
By combining Eqs.\rf{bar-g1},
\rf{integr-const1} and \rf{Omega-eqN}, the relation between ${\bar g}_{\m\n}$ and $g_{\m\n}$ can be written as
\be
{\bar g}_{\m\n} = \chi_1\O\,g_{\m\n} \; .
\lab{bar-g-2}
\ee

In this way, the Eq.\rf{einstein-like-eqs1} 
for the rescaled  metric ${\bar g}_{\m\n}$ \rf{bar-g-2}, 
\textsl{i.e.}, the Einstein-frame can be rewritten as 
\be
R_{\m\n}({\bar g}) - \h {\bar g}_{\m\n} R({\bar g}) = \h T^{\rm eff}_{\m\n},
\lab{eff-einstein-eqs}
\ee
where the effective energy-momentum tensor in the Einstein-frame becomes 
\be
T^{\rm eff}_{\m\n} = g_{\m\n} L_{\rm eff} - 2 \partder{}{g^{\m\n}} L_{\rm eff} ,
\lab{EM-tensor-eff1}
\ee
in which the effective Einstein-frame scalar fields Lagrangian results
\be
L_{\rm eff}(\vp_1,\vp_2) = \frac{1}{\chi_1\O}\Bigl\{ L^{(1)} + M_1 +
\frac{\chi_2}{\chi_1\O}\Bigl\lb L^{(2)} + M_2 + 
\eps (L^{(1)} + M_1)^2\Bigr\rb\Bigr\} \; .
\lab{L-effN}
\ee

For the effective Lagrangian  $L_{\rm eff}$ in terms of the Einstein-frame
metric ${\bar g}_{\m\n}$ \rf{bar-g-2} we can consider the short-hand notation for the
scalar kinetic terms $X_1$ and $X_2$ such that
\be
X_1 \equiv - \h {\bar g}^{\m\n}\pa_\m \vp_1 \pa_\n \vp_1,\,\,\,\,\mbox{and}\,\,\,\,X_2 \equiv - \h {\bar g}^{\m\n}\pa_\m \vp_2 \pa_\n \vp_2,
\lab{X-def}
\ee
and the two Lagrangians  $L^{(1,2)}$ become 
\be
L^{(1)} = \chi_1\O\, \left[X_1+X_2\right] - V \quad ,\quad L^{(2)} = \chi_1\O\,\left[b_1 e^{-\a_1\vp_1}X_1+b_2 e^{-\a_2\vp_2}X_2\right] + U \; ,
\lab{L-1-2-Omega}
\ee
with $V$ and $U$ given by  Eq.\rf{U,V}.

From Eqs.\rf{chi-11} and \rf{Omega-eqN}, taking into account \rf{L-11}, 
we get
\be
\frac{1}{\chi_1\O} = \frac{(V-M_1)}{2\chi_2\Bigl\lb U+M_2 + \eps (V-M_1)^2\Bigr\rb}
\,\left[ 1 - \chi_2 \left[\left(\frac{b_1 e^{-\a_1\vp_1}}{V-M_1}-2\eps\right) X_1+\Bigl(\frac{b_2 e^{-\a_2\vp_2}}{V-M_1} - 2\eps\Bigr) X_2\right] \right]\; .
\lab{chi-OmegaN}
\ee
Upon substituting expression \rf{chi-OmegaN} into \rf{L-effN} we arrive at the
explicit form for the Einstein-frame scalar Lagrangian:
\be
L_{\rm eff} = A_1(\vp_1,\vp_2 ) X_1 +A_2(\vp_1,\vp_2) X_2 + B_1(\vp_1, \vp_2) X_1^2 +B_2(\vp_1, \vp_2) X_2^2+B_{12}(\vp_1, \vp_2)X_1X_2  - U_{\rm eff}(\vp_1, \vp_1 ) \; ,
\lab{L-eff-final}
\ee
where the functions $A_1(\vp_1,\vp_2 )$ and $A_2(\vp_1,\vp_2 )$ are given by
$$
A_1(\vp_1,\vp_2 ) = 1 + \Bigl\lb \h b_1 e^{-\a_1\vp_1} - \eps (V - M_1)\Bigr\rb
\frac{V - M_1}{U + M_2 + \eps (V - M_1)^2}= 1 +
$$
\be
 \Bigl\lb \h b_1 e^{-\a_1\vp_1} - \eps\(f_1 e^{-\a_1\vp_1} +g_1 e^{-\a_2\vp_2} - M_1\) \Bigr\rb
\,\frac{f_1 e^{-\a_1\vp_1} + g_1 e^{-\a_2\vp_2}- M_1}{f_2 e^{-2\a\vp_1} + g_2 e^{-2\a\vp_2}+ M_2 + \eps (f_1 e^{-\a\vp_1} + g_1 e^{-\a_2\vp_2}  - M_1)^2}
\; , 
\ee
and 
$$
A_2(\vp_1,\vp_2 ) = 1 + \Bigl\lb \h b_2 e^{-\a_2\vp_2} - \eps (V - M_1)\Bigr\rb
\frac{V - M_1}{U + M_2 + \eps (V - M_1)^2}
$$
\be
= 1 + \Bigl\lb \h b_2 e^{-\a_2\vp_2} - \eps\(f_1 e^{-\a\vp_1} +g_1 e^{-\a_2\vp_2} - M_1\) \Bigr\rb
\,\frac{f_1 e^{-\a_1\vp_1} + g_1 e^{-\a_2\vp_2}- M_1}{f_2 e^{-2\a\vp_1} + g_2 e^{-2\a\vp_2}+ M_2 + \eps (f_1 e^{-\a\vp_1} + g_1 e^{-\a_2\vp_2}  - M_1)^2}.
\lab{A-def}
\ee
The coefficient $B_1(\vp_1,\vp_2 )$ is defined as
\br
B_1(\vp_1,\vp_2 ) = \chi_2 \frac{\eps\Bigl\lb U + M_2 + (V - M_1) b_1 e^{-\a_1\vp_1}\Bigr\rb
- \frac{1}{4} b_1^2 e^{-2\a\vp_1}}{U + M_2 + \eps (V - M_1)^2}
\nonu\\
= \chi_2 \frac{\eps\Bigl\lb f_2 e^{-2\a\vp_1}+g_2 e^{-2\a\vp_2} + M_2 +
(f_1 e^{-\a\vp_1} + g_1e^{-\a_2\vp_2} - M_1)b_1 e^{-\a_1\vp_1}\Bigr\rb -\frac{1}{4} b_1^2 e^{-2\a_1\vp_1}
}{f_2 e^{-2\a_1\vp_1} + g_2 e^{-2\a_2\vp_2} +M_2 + \eps (f_1 e^{-\a_1\vp_1}+g_2 e^{-\a_2\vp_2} - M_1)^2} \; ,
\lab{B-def}
\er
and for the function $B_2(\vp_1,\vp_2)$ we obtain
\br
B_2(\vp_1,\vp_2 ) = \chi_2 \frac{\eps\Bigl\lb U + M_2 + (V - M_1) b_2 e^{-\a_2\vp_2}\Bigr\rb
- \frac{1}{4} b_2^2 e^{-2\a\vp_2}}{U + M_2 + \eps (V - M_1)^2}
\nonu\\
= \chi_2 \frac{\eps\Bigl\lb f_2 e^{-2\a\vp_1}+g_2 e^{-2\a\vp_2} + M_2 +
(f_1 e^{-\a\vp_1} + g_1e^{-\a_2\vp_2} - M_1)b_2 e^{-\a_2\vp_2}\Bigr\rb -\frac{1}{4} b_2^2 e^{-2\a_2\vp_2}
}{f_2 e^{-2\a_1\vp_1} + g_2 e^{-2\a_2\vp_2} +M_2 + \eps (f_1 e^{-\a_1\vp_1}+g_2 e^{-\a_2\vp_2} - M_1)^2} \; ,
\lab{B-def1}
\er
and the coefficient $B_{12}(\vp_1,\vp_2)$  becomes
$$
B_{12}(\vp_1,\vp_2)=\chi_2E_0\Big[-E_1b_2e^{-\a_2\vp_2}-E_2b_1e^{-\a_1\vp_1}+2E_0E_1E_2[(M_2+U)+
$$
\be
\epsilon(M_1-V)^2]+2\epsilon[(1/E_0)-(E_1+E_2)(M_1-V)]\Big],
\ee
where the quantities $E_0$, $E_1$ and $E_2$ are defined as
$$
E_0=\frac{(V-M_1)}{2\chi_2[U+M_2+\epsilon(V-M_1)^2]},\,\,\,E_1=\chi_2\left[\frac{b_1e^{-\a_1\vp_1}}{V-M_1}-2\epsilon\right],\,\,\,\mbox{and}\,\,\,\,E_2=\chi_2\left[\frac{b_2e^{-\a_2\vp_2}}{V-M_1}-2\epsilon\right].
$$

The effective scalar field potential as a function of the scalar fields yields
$$
U_{\rm eff} (\vp_1,\vp_2) = 
\frac{(V - M_1)^2}{4\chi_2 \Bigl\lb U + M_2 + \eps (V - M_1)^2\Bigr\rb}=
$$
\be
 \frac{(f_1 e^{-\a_1\vp_1}+g_1 e^{-\a_2\vp_2} -M_1)^2}{4\chi_2\,\Bigl\lb 
f_2 e^{-2\a_1\vp_1} +g_2 e^{-2\a_2\vp_2} + M_2 + \eps (f_1 e^{-\a_1\vp_1} + g_1 e^{-\a_2\vp_2}-M_1)^2\Bigr\rb} \; ,
\lab{U-eff}
\ee
where we have used for $V$ and $U$, the expressions given by  Eq.\rf{U,V}.

\section{Three Flat Regions from an Effective Scalar Potential}
\label{flat-regions}

The crucial feature of $U_{\rm eff} (\vp_1,\vp_2)$ is the presence of three infinitely large
flat regions. In this sense, we have one for large positive values of the scalar fields  $\vp_1$ and  $\vp_2$ and two others for the limits $\vp_1 \rightarrow -\infty$ and $\vp_2 \rightarrow -\infty$.

For large negative values of $\vp_1$, which we will choose to describe the very early phase of the universe, meaning the emergent phase and inflation we have for the effective potential and the
coefficient functions in the Einstein-frame scalar Lagrangian 
\rf{L-eff-final}-\rf{U-eff}:
\be
U_{\rm eff}(\vp_1, \vp_2 ) \simeq U_{\rm eff}{(-\infty,\varphi_2)} =U_{eff} =
\frac{f_1^2/f_2}{4\chi_2 (1+\eps f_1^2/f_2)} \; ,
\lab{U-minus1} \\
\ee
\be
A_1(\vp_1,\vp_2) \simeq A_1{(-\infty,\varphi_2)} =A_1= \frac{1+\h b_1 f_1/f_2}{1+\eps f^2_1/f_2} \;\; ,\;\; 
B_1(\vp_1,\vp_2) \simeq B_1{(-\infty,\varphi_2)} =B_1=
- \chi_2 \frac{b_1^2/4f_2 - \eps (1+ b_1 f_1/f_2)}{1+\eps f^2_1/f_2} \; .
\lab{A-B-minus1}
\ee
For the terms $A_2$ and $B_2$ in the limit in which $\vp_1 \rightarrow -\infty$ we have
\be
A_2(\vp_1,\vp_2) \simeq A_2{(-\infty,\varphi_2)} =A_2=\frac{1}{1+\eps f^2_1/f_2},\,\,\,\mbox{and}\,\,\,\,\, B_2(\vp_1,\vp_2) \simeq B_2{(-\infty,\varphi_2)} =B_2= \frac{\chi_2  \eps}{1+\eps f^2_1/f_2} .
\ee

For the coefficient $B_{12}(\vp_1,\vp_2) $ in the limit in which the scalar field $\vp_1 \rightarrow -\infty$ becomes
\be
B_{12}(\vp_1,\vp_2) \simeq B_{12}{(-\infty,\varphi_2)} =B_{12}=
\chi_2\tilde{E}_0\Big[-\tilde{E}_2b_1+\tilde{E}_0f_2+\frac{2\epsilon}{\tilde{E}_0}+\epsilon\tilde{E}_0f_1^2+2\epsilon f_1(\tilde{E}_1+\tilde{E}_2)\Big],
\lab{A-B-minus1a}
\ee
where
$$
\tilde{E}_0=\frac{f_1}{2\chi_2[f_2+\epsilon f_1^2]},\,\,\,\,\tilde{E}_1=\chi_2\left[\frac{b_1}{f_1}-2\epsilon\right],\,\,\mbox{and}\,\,\,\,\tilde{E}_2=-2\chi_2\epsilon.
$$

For the second flat region we will consider that the scalar field  $\vp_2 \rightarrow -\infty$ such that the effective potential in the second flat region is given by
\be
U_{\rm eff}(\vp_1, \vp_2 ) \simeq U_{\rm eff}{(\varphi_1,-\infty)} = U_{\rm eff\,g}=
\frac{g_1^2/g_2}{4\chi_2 (1+\eps g_1^2/g_2)} \; ,
\lab{U-minus2} \\
\ee
and the kinetic coefficients   $A_2$ and $B_2$ in the limit in which $\vp_2 \rightarrow -\infty$ result
\be
 A_2{(\vp_1,-\infty)} =A_{2g}= \frac{1+\h b_2 g_1/g_2}{1+\eps g^2_1/g_2}  ,\;\;
 B_2{(\vp_1,-\infty)} = B_{2g}=
- \chi_2 \frac{b_2^2/4g_2 - \eps (1+ b_2 g_1/g_2)}{1+\eps g^2_1/g_2} \; ,
\lab{A-B-minus2}
\ee
and the terms $A_1$ and $B_1$ in this limit result
\be
A_1(\vp_1,\vp_2) \simeq A_1{(\vp_1,-\infty)} =A_{1g}= \frac{1}{1+\eps g^2_1/g_2} \;\; ,\;\; 
B_1(\vp_1,\vp_2) \simeq B_1{(\vp_1,-\infty)} = B_{1g}=
 \frac{\chi_2 \eps}{1+\eps g^2_1/g_2} \; .
\lab{A-B-minus2b}
\ee

For the coefficient $B_{12}(\vp_1,\vp_2) $ in the limit in which the scalar field $\vp_2 \rightarrow -\infty$ we have 
\be
B_{12}(\vp_1,\vp_2) \simeq B_{12}{(\varphi_1,-\infty)} =B_{12g}=
\chi_2\tilde{E}_3\Big[-\tilde{E}_2b_2+\tilde{E}_3g_2+\frac{2\epsilon}{\tilde{E}_3}+\epsilon\tilde{E}_3g_1^2+2\epsilon g_1(\tilde{E}_4+\tilde{E}_2)\Big],\label{B12}
\ee
where
$$
\tilde{E}_3=\frac{g_1}{2\chi_2[g_2+\epsilon g_1^2]},\,\,\,\mbox{and}\,\,\,\,\,\tilde{E}_4=\chi_2\left[\frac{b_2}{g_1}-2\epsilon\right].
$$

In the third flat region for large positive $\vp_1$ and also $\vp_2$, we find that the effective  potential reduces to 
\be
U_{\rm eff}(\vp_1,\vp_2) \simeq U_{\rm eff}{(+\infty,+\infty)} = U_{\rm eff (+)}=
\frac{M_1^2/M_2}{4\chi_2 (1+\eps M_1^2/M_2)} \; ,
\lab{U-M} \\
\ee
and the kinetic coefficients are
\be
A_1(\vp_1,\vp_2)= A_2(\vp_1,\vp_2)\simeq A_{(+)} \equiv \frac{M_2}{M_2 + \eps M_1^2} \quad ,\quad
B_1(\vp_1,\vp_2)= B_2(\vp_1,\vp_2)\simeq B_{(+)} \equiv \eps\chi_2 \frac{M_2}{M_2 + \eps M_1^2} \; ,
\lab{A-B-M}
\ee
and for this limit we find that the coefficient $B_{12}(\vp_1,\vp_2)$ becomes
\be
B_{12}(\vp_1,\vp_2)= B_{12}(+\infty,+\infty)=B_{12(+)}=2\eps\chi_2 \frac{M_2}{M_2 + \eps M_1^2}=2\,B_{(+)}.
\lab{B12-M}
\ee

We will consider that the flat region \rf{U-minus1} corresponds 
to the evolution of the early  universe (emergent and inflation). On the other hand, the flat regions \rf{U-minus2} and  \rf{U-M} concern 
to the evolution of the late universe with a two phase structure.

In particular, if we assume the order of magnitude of the coupling parameters 
in the effective potential given by Eq.\rf{U-minus1}, are 
$f_1 \sim f_2 \sim (10^{-2} M_{Pl})^4$, then
 the order of
magnitude of the vacuum energy density of the early universe during the inflationary epoch yields
\be
U_{\rm eff}{(-\infty,\varphi_2)} =U_{eff}\sim f_1^2/f_2 \sim 10^{-8} M_{Pl}^4 \; ,
\lab{U-minus-magnitude}
\ee
here we have assumed that the parameter $\epsilon$ is small and the integration constant $\chi_2\sim \mathcal{O}(1)$.

In order to study the evolution of the universe from an emergent and inflationary scenarios to dark epoch, we consider that
 the standard Friedman-Lemaitre-Robertson-Walker
space-time metric is given by 
\be
ds^2 = - dt^2 + a^2(t) \Bigl\lb \frac{dr^2}{1-K r^2}
+ r^2 (d\th^2 + \sin^2\th d\phi^2)\Bigr\rb,
\lab{FLRW}
\ee
where $a(t)$ denotes the scale factor and $K$ corresponds to the space curvature.
 
 By assuming that the matter is described by a perfect fluid with an energy density and pressure $\rho$ and $p$, we have that 
 the associated Friedmann equations are
\be
\frac{\addot}{a}= - \frac{1}{12} (\rho + 3p)  ,\quad
H^2 + \frac{K}{a^2} = \frac{1}{6}\rho  ,\;\;  \; \,\,\,\mbox{and}\,\,\,\,\;\,\;\dot{\rho}+3H(\rho+p)=0,
\lab{friedman-eqs}
\ee
where $H= \frac{\adot}{a}$ is the Hubble parameter. Also, 
here the energy density and pressure associated to  the scalar fields $\vp_1 = \vp_1 (t)$ and $\vp_2 = \vp_2 (t)$   are defined as 
\be
\rho =  A_1(\vp_1, \vp_2) X_1 + A_2(\vp_1, \vp_2) X_2 + 3 B_1(\vp_1, \vp_2) X_1^2 + 3 B_2(\vp_1, \vp_2) X_2 ^2 + 3 B_{12}(\vp_1, \vp_2) X_1X_2+U_{\rm eff}(\vp _1,\vp_2) \; ,
\lab{rho-def} 
\ee
and
\be
p =  A_1(\vp_1, \vp_2) X_1 + A_2(\vp_1, \vp_2) X_2 +  B_1(\vp_1, \vp_2) X_1^2 +  B_2(\vp_1, \vp_2) X_2 ^2 +B_{12}(\vp_1, \vp_2) X_1X_2- U_{\rm eff}(\vp_1,\vp_2) \; .
\lab{p-def}
\ee

Henceforth the dots indicate derivatives with respect to the time $t$ and we have assumed that the scalar fields are homogeneous. 

\section{Non-Singular Emergent Universe}
\label{cuatro}

In this section we will exhibit 
 that under specific conditions  on the parameters-space 
 there
exist an era  preceding the inflationary  epoch. In this sense, we obtain a  
cosmological solution of the Einstein-frame system
in our model from Lagrangian \rf{L-eff-final} describing 
a non-singular ``emergent universe'' \ct{emerg,emerg2} 
when one the scalar field $\varphi_1$ evolves on the first flat region for large negative,    where the effective potential is described by Eq.\rf{U,V}. We mention that 
for previous analysis    of ``emergent universe'' epoch considering  modified-measure gravity-matter theories
with one non-Riemannian and one standard Riemannian integration measures but with one scalar field was studied in \ct{ourquintessence}, see also
Refs.\ct{TMT-recent-1-a,TMT-recent-1-c}.

The emergent universe,  
is characterized  through the standard 
Friedmann-Lemaitre-Robertson-Walker space-time metric \rf{FLRW} as a solution
of Eq.\rf{friedman-eqs} under  the condition on the Hubble parameter $H=0$ such that
\be
 a(t) = a_0 = {\rm const} \,,\;\; \rho + 3p =0 \quad ,\quad \mbox{and}\,\,\,\,
\frac{K}{a_0^2} = \frac{1}{6}\rho ~(= {\rm const}) \; ,
\lab{emergent-cond}
\ee
with $\rho$ and $p$ given by Eqs. \rf{rho-def}-\rf{p-def}. 
From Eq.\rf{emergent-cond}  and noticing that the combination $\rho - 3p= -2(A_1(\vp_1, \vp_2) X_1 + A_2(\vp_1, \vp_2) X_2) + 4U_{\rm eff}(\vp _1,\vp_2) $ does not contain the non linear terms and since   $\rho - 3p =2\rho-(\rho + 3p) = \frac{12K}{a_0^2}$, then we can  find that the relation between $X_{10}$ and $X_{20}$, 
is given by
\be
X_{20}=\frac{\dot{\vp}_{20}^2}{2}=\frac{C_0-A_1\,X_{10}}{A_2}>0,\label{X20}
\ee
where $ C_0 = 2U_{eff}- \frac{6K}{a_0^2} $. Here we note that the parameter $C_0>0$, since $A_1$ and $A_2$ are positive and $C_0>A_1X_{10}$.
In this sense, in multi-fields model the spatial curvature of the universe cannot be too big, as opposed to the single field  TMT  model studied before in \cite{Gravityassisted}, where no such restriction was found.

Using this we now find that the emergent universe condition \rf{emergent-cond} implies that the   $\vp_1$-velocity
$\vpdot_1 \equiv \vpdot_{10}$ and $\vp_2$-velocity
$\vpdot_2 \equiv \vpdot_{20}$
are time-independent and the kinetic term  $X_{10}=\dot{\vp}_{10}^2/2$ satisfies the  
algebraic equation:
\be
X_{10}^2[B_1+B_2A_1^2/A_2^2-B_{12}A_1/A_2]+X_{10}[-6B_2A_1C_0/A_2^2+B_{12}C_0/A_2]+[2C_0/3+B_2^2C_0^2/A_2^2-U_{eff}/3]=0.
\lab{vpdot-eq}
\ee

Defining the quantities $F_1$,$F_2$ and $F_3$ as $ F_1 = B_1+B_2A_1^2/A_2^2-B_{12}A_1/A_2 $,  $ F_2 =-6B_2A_1C_0/A_2^2+B_{12}C_0/A_2$ and  
$ F_3 = 2C_0/3+B_2^2C_0^2/A_2^2-U_{eff}/3 $, 
the solution of the kinetic term $X_{01}$ yields
\be
X_{10} =  \frac{1}{2 F_1 }\Bigr\lb - F_2 \pm
\sqrt{ F_2 ^2 - 4  F_1  F_3  } \Bigr\rb =\frac{1}{2}\left[-\frac{F_2}{F_1}\pm \sqrt{\left(\frac{F_2}{F_1}\right)^2-\frac{4F_3}{F_1}}\right]>0\; .
\lab{vpdot-sol}
\ee
Here the ratio $F_2/F_1$ becomes
$$
\frac{F_2}{F_1}=-C_0\,\left(\frac{-A_2B_{12}+6A_1B_2}{A_2^2B_1-A_1A_2B_{12}+A_1^2B_2}\right),
$$
and assuming that the parameter $\epsilon \ll 1$, then we have that the coefficients  $A_1>0$, $A_2>0$, $B_1<0$, $B_2\sim 0$ and $B_{12}>0$
with which the ratio $F_2/F_1$ is negative. Also, we find that the quantity 
$$
\left(\frac{F_2}{F_1}\right)^2-\frac{4F_3}{F_1}=\frac{4(A_2^2B_1-A_1A_2B_{12}+A_1^2B_2)[A_2^2(12k/a_0^2-3U_{eff})-3B_2^2C_0^2]+3(A_2B_{12}-6A_1B_2)^2C_0^2}{3A_2^4\left(B_1+\frac{A_1(A_1B_2-A_2B_{12})}{A_2^2}\right)^2}.
$$
As before considering that the parameter $\epsilon \ll 1$, we find the the quantity $(F_2/F_1)^2-4F_3/F_1$ is positive. Here we have considered that $3U_{eff}/2>6k/a_0^2$ i.e., $C_0>2k/a_0^2$. Also, we note that the term $F_3/F_1$ is a negative quantity for $\epsilon \ll 1$, then  in the following we will  consider the positive sign of Eq.\rf{vpdot-sol}. Thus, we require that these solutions for $X_{10}$ and $X_{20}$ be reals and positives.

In order to study the stability of this solution,   we can  analyze  the perturbation associated to the energy density $\delta\rho$ in which 
\be
\frac{\d \addot}{a_0} + \frac{1}{12} (\d\rho + 3 \d p)=0 \,\,\, ,\quad \mbox{with}\,\,\,\,\,\,\,
\d \rho = - \frac{2\rho_0}{a_0} \d a ,\,
\lab{density pert}
\ee
where we need to express $\d \rho$  in terms of $\d \vp_1$  and $\d \vp_2$.
Notice that due to the shift symmetries that exist in the flat region,
$\vp_1 \rightarrow \vp_1 + constant $ and $\vp_2 \rightarrow \vp_2 + constant $,
we have conserved quantities that should be preserved in these perturbations, and this ends up providing us with an important relation between   $\d \vp_1$  and $\d \vp_2$.
The conservation law related to the symmetry
$\vp_1 \rightarrow \vp_1 + constant $ is
\be
C_1=a^3\left[A_1\dot{\vp}_1+B_{12}\dot{\vp}_1 X_2+2B_1\dot{\vp}_1X_1\right] ,\label{C1}
\ee
and the conservation law associated to the   symmetry
$\vp_2 \rightarrow \vp_2 + constant $  results
\be
C_2=a^3\left[A_2\dot{\vp}_2+B_{12}\dot{\vp}_2 X_1+2B_2\dot{\vp}_2X_2\right] ,\label{C2}
\ee
where $C_1$ and $C_2$ are two arbitrary constants.
Also, the quantities $a_0$ and $\rho_0$ are defined as
\be
a_0^2 = \frac{6K}{\rho_0} \quad ,\quad
\rho_0 =A_1 X_{10} + A_2 X_{20} + 3 B_1 X_{10}^2 + 3 B_2 X_{20} ^2 + 3 B_{12} X_{10}X_{20} +U_{\rm eff}(-\infty,\vp_2)  \; ,
\lab{emergent-univ}
\ee
with $X_{10}$ and $X_{20}$ given by  Eqs.(\ref{X20}) and \rf{vpdot-sol}.

On the other hand, we perturb Friedmann  equations  given by \rf{friedman-eqs}
and the expressions for $\rho,\, p$ \rf{rho-def}-\rf{p-def} w.r.t.  
$a(t) = a_0 + \d a (t)$ and $\vpdot_1 (t) = \vpdot_{10} + \d \vpdot_1 (t)$, $\vpdot_2 (t) = \vpdot_{20} + \d \vpdot_2 (t)$ but
keep the effective potential on the flat region in which $\vp_1\rightarrow-\infty$,  $U_{\rm eff} = U(-\infty,\vp_2)$:
\be
\frac{\d \addot}{a_0} + \frac{1}{12} (\d\rho + 3 \d p)=0 \quad ,\quad
\d \rho = - \frac{2\rho_0}{a_0} \d a \; ,
\lab{delta-friedman} 
\ee
\be
\d\rho =\tilde{A}\,\d\vpdot_{1}+\tilde{B}\,\d\vpdot_{2}=
- \frac{2\rho_0}{a_0} \d a,
\ee
where the quantities $\tilde{A}$ and $\tilde{B}$ are given by
\be
\tilde{A} = \(A_1\vpdot_{10} + 3 B_{1}\vpdot_{10}^3
+\frac{3}{2}B_{12}\vpdot_{10}\vpdot_{20}^2\),\,\,\,\mbox{and}\,\,\,
\tilde{B}= \left(A_2\vpdot_{20} + 3 B_{2}\vpdot_{20}^3
+\frac{3}{2}B_{12}\vpdot_{20}\vpdot_{10}^2\right)
 \;\; ,\;\;
\ee
and  for the perturbation associated to pressure we have
\be
\d p = \tilde{C}
\d\vpdot_{1}+ \tilde{D}
\d\vpdot_{2} ,
\lab{delta-rho-p}
\ee
with $\tilde{C}$ and $\tilde{D}$ are defined as
\be
\tilde{C}=\(A_1\vpdot_{10} +  B_{1}\vpdot_{10}^3
+\frac{1}{2}B_{12}\vpdot_{10}\vpdot_{20}^2\),\,\,\,\mbox{and}\,\,\,\,\tilde{D}=\left(A_2\vpdot_{20} +  B_{2}\vpdot_{20}^3
+\frac{1}{2}B_{12}\vpdot_{20}\vpdot_{10}^2\right).
\ee
Also, the perturbation equation for the scalar fields from Eq.(\ref{C1}) becomes
\be
\d\vpdot_{2}=D_0\d\vpdot_{1}+D_1\delta a,
\label{Pe1}
\ee
where the coefficients $D_3$ and $D_4$ are defined
$$
D_0=-\Big[\frac{A_1+B_{12}X_{20}+6B_1 X_{10}}{B_{12}\vpdot_{10}\vpdot_{20}}\Big]
,\,\,\,\,\,\mbox{and}\,\,\,\,\,\,D_1=-\frac{3}{a_0}\Big[\frac{A_1\vpdot_{10}+B_{12}\vpdot_{10}X_{20}+B_1\vpdot_{10}X_{10}}{B_{12}\vpdot_{10}\vpdot_{20}}\Big].
$$
However, from Eq.(\ref{C2}) we have
\be
\d\vpdot_{2}=D_2\d\vpdot_{1}+D_3\delta a,
\label{Pe2}
\ee
where the terms $D_2$ and $D_3$ are given by
$$
D_2=-\Big[\frac{B_{12}\vpdot_{10}\vpdot_{20}}{A_2+B_{12}X_{10}+6B_2 X_{20}}\Big]
,\,\,\,\,\,\mbox{and}\,\,\,\,\,\,D_3=-\frac{3}{a_0}\Big[\frac{A_2\vpdot_{20}+B_{12}\vpdot_{20}X_{10}+B_2\vpdot_{20}X_{20}}{A_2+B_{12}X_{10}+6B_2 X_{20}}\Big].
$$
In order to have consistency between Eqs.(\ref{Pe1}) and (\ref{Pe2}) we must have $D_0=D_2$ and $D_1=D_3$, respectively.  

By expressing $\d\vpdot_1$ and $\d\vpdot_2$ in terms of the  $\d a$ and
substituting into the first Eq.\rf{delta-friedman} we get a harmonic
oscillator type equation for $\d a$:
\be
\d \addot + \om^2 \d a = 0 \quad ,\quad
\om^2=\frac{1}{12(\tilde{A}+\tilde{B}D_0)} \left[(\tilde{B}a_0D_1-2\rho_0)(\tilde{A}+3\tilde{C})+a_0D_1\tilde{A}+2D_0\tilde{B}a_0D_1-2D_0\rho_0\right] \; .
\lab{stability-eq}
\ee
To obtain the stability of the emergent universe solution
we need to consider the parameter space for which $\om^2>0$. Here we have that $(\tilde{A}+\tilde{B}D_0)\neq 0$.

\begin{figure}
 	\centering
\includegraphics[width=0.68\textwidth]{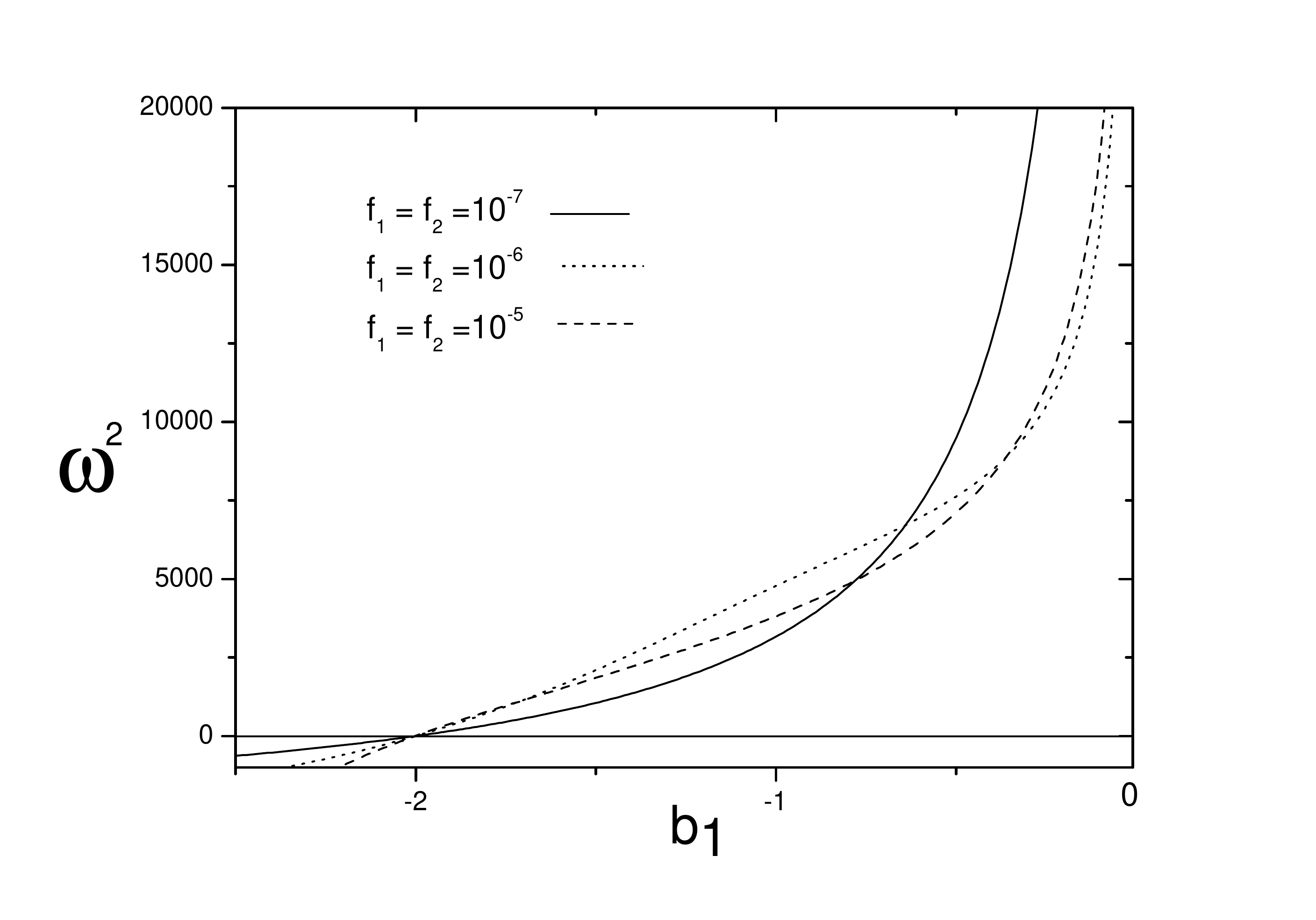}
\caption{\it{ {The figure depicts the frequency square as a function of the parameter $b_1$, for different values of $f_1$ and $f_2$. Here we have fixed the values $\chi_2=1$, $\epsilon=10^{-4}$ (in units of $M_{Pl}^{-2}$) and $\rho_0=(3/2)U_{\rm eff}$. } }}

 	\label{fig1}
\end{figure}

Because of the complexity of the above expression for $\omega^2$, we perform  a numerical analysis of the  dependence of $\omega^2$ versus the parameter $b_1$ which  has been plotted in  Fig.\ref{fig1}.
In this context, in  Fig.\ref{fig1} we show the dependence  of the frequency square in terms of the parameter $b_1$, for various values of the parameters $f_1$ and $f_2$. In order to satisfy  the stable  emergent universe solution, the parameter $b_1$ is negative and it has to be in the interval $-2<b_1<0$. Here we observe that this range for the parameter $b_1$
does not depend of the values of $f_1$ and $f_2$.

\section{ Inflationary epoch}\label{inflacion}

In this section we will analyze the inflationary universe in the situation in which the scalar field $\vp_1\rightarrow-\infty$ where the effective potential corresponds to first flat region and the kinetic coefficients   are given by Eqs.\rf{A-B-minus1}-\rf{A-B-minus1a}.

The full equations for the scalar fields $\vp_1$ and $\vp_2$ are given by
$$
\ddot{\vp}_1\left(A_1+6B_1X_1+B_{12}X_2\right)+3H\dot{\vp}_1\left[A_1+2B_1X_1+B_{12}X_2+\frac{B_{12}}{3H}\dot{\vp_2}\ddot{\vp}_2\right]
$$
\be
=-U_{eff,\vp_1}+A_{1,\vp_1}X_1+A_{2,\vp_1}X_2+B_{1,\vp_1}X_1^2+B_{2,\vp_1}X_2^2+B_{12,\vp_1}X_1\,X_2,\label{fa1}
\ee
and 
$$
\ddot{\vp}_2\left(A_2+6B_2X_2+B_{12}X_1\right)+3H\dot{\vp}_2\left[A_2+2B_2X_2+B_{12}X_1+\frac{B_{12}}{3H}\dot{\vp_1}\ddot{\vp}_1\right]
$$
\be
=-U_{eff,\vp_2}+A_{2,\vp_2}X_2+A_{1,\vp_2}X_1+B_{2,\vp_2}X_1^2+B_{1,\vp_2}X_1^2+B_{12,\vp_2}X_1\,X_2,\label{fa2}
\ee
where the notation $,\vp_1$ and $,\vp_2$ denote $\partial /\partial\vp_1$ and  $\partial /\partial\vp_2$, respectively.

Introducing  the  ``slow-roll'' parameters for the scalar fields  \ct{slow-roll-param}
\be
\vareps_{H} = - \frac{\Hdot}{H^2} , \quad 
\eta_1 = -\frac{\vpddot_1}{H\vpdot_1} \; ,\,\,\,\,\,\mbox{and}\,\,\,\,\,\,\eta_2 = -\frac{\vpddot_2}{H\vpdot_2},
\label{SLR-def1}
\ee
and assuming that these parameters are $\vareps_{H}$, $\eta_1$ and $\eta_2\ll 1$, then  we can ignore the terms $\ddot{\vp}_{1,2}$ and non-linear, such that the Eqs.(\ref{fa1}) and (\ref{fa2}) together with the Friedmann equation reduce to 
\be
3A_1H\vpdot_1 + U_{ eff,\vp_1} \simeq 0  ,\,\,\,\,\,\,3A_2H\vpdot_2 + U_{ eff,\vp_2} \simeq 0,\,\,\,\mbox{and}\,\quad H^2 \simeq \frac{1}{6} U_{\rm eff} \; .
\label{72}
\ee
From Eq.\rf{U-eff} and considering  the region in which $f_1e^{-\a_1\vp_1}+g_1e^{-\a_2\vp_2}\gg M_1$ and $f_2e^{-2\a_1\vp_1}+g_2e^{-2\a_2\vp_2}\gg M_2$, the effective potential becomes
\be
U_{eff}(\vp_1,\vp_2)=\frac{(f_1e^{-\a_1\,\vp_1}+g_1e^{-\a_2\vp_2})^2}{4\chi_2(f_2e^{-2\a_1\vp_1}+g_2e^{-2\a_2\vp_2}+\eps (f_1 e^{-\a_1\vp_1} + g_1 e^{-\a_2\vp_2})^2)}.\lab{U4}
\ee
Note that from the above expression  we have that the effective potential  can be rewritten in term of a  single scalar field $\hat{\phi}$ in which  
\be
U_{eff}(\hat\phi)=\frac{(f_1e^{-\sqrt{\a_1^2+\a_2^2}\hat{\phi}}+g_1)^2}{4\chi_2(f_2e^{-2\sqrt{\a_1^2+\a_2^2}}\hat{\phi}+g_2 + \eps (f_1 e^{-\sqrt{\a_1^2+\a_2^2}\hat{\phi}} + g_1 )^2) },\lab{UUU}
\ee
where $$\hat{\phi}=  \frac{\a_1\vp_1-\a_2\vp_2}{\sqrt{\a_1^2+\a_2^2} }.$$
For large negative value of the scalar field $\vp_1$, which means also that $\hat{\phi}$ is also large and negative  we have that we can neglect $ g_1$ and $ g_2$ in the denominator and keeping the leading contribution beyond the constant 
$\frac{f_1^2}{4\chi_2(f_2+\epsilon f_1^2)}$ of the infinite plateau,  we get that the effective potential becomes (going back to the $\vp_1, \vp_2$  variables)


\be
U_{ eff} \simeq \left(\frac{f_1^2+2f_1g_1e^{\sqrt{\a_1^2+\a_2^2}\hat{\phi}}}{4\chi_2(f_2+\epsilon f_1^2)}\right)=
\left(\frac{f_1^2+2f_1g_1e^{\a_1\vp_1-\a_2\vp_2}}{4\chi_2(f_2+\epsilon f_1^2)}\right) \; .
\lab{Usowroll}
\ee
The potential \rf{UUU}
interpolates between the vacuums with asymptotic values $\frac{f_1^2}{4\chi_2(f_2+\epsilon f_1^2)}$ and $\frac{g_1^2}{4\chi_2(g_2+\epsilon g_1^2)}$ independent of the choice of the relative signs of $g_1$, and $f_1$. Beyond, this the more detailed  structure of the potential is different if the relative sign is the same or if it is opposite.
In the case  signs of $g_1$, and $f_1$ are the same, the potential does not have a minimum where the potential is zero, and although the potential still connects the two vacuum, the potential shows a bump (as it is apparent also by the fact that the $f_1g_1$ term in \rf{Usowroll} raises  the energy density over the energy density of the plateau ) that pushes slow roll solutions  in the opposite to the desired direction, i.e., not in the direction necessary to interpolate with the desired vacuum, If the case the signs of $g_1$, and $f_1$ are the opposite is considered,  the potential does have a minimum where the potential is  zero, and now the potential does not show a bump and the slow roll solutions causes the scalar field to move  to the desired direction, i.e., in the direction necessary to interpolate with the desired vacuum. We can choose therefore the case where the signs of $g_1$, and $f_1$ are the opposite. In the following, we will choose the parameter $f_1>0$ and then the parameter $g_1$ is negative.

Now, in order to study the inflationary epoch, we will consider the slow roll approximation. In this framework, 
from Eq.(\ref{72}) we find that the relation between the scalar field $\vp_1$ and $\vp_2$, under the slow roll approximation yields 
\be
\vp_1=-\left(\frac{A_2\a_1}{A_1\a_2}\right)\vp_2+C,
\label{FF1}
\ee
where $C$ denotes an integration constant and here we have used the effective potential given by Eq.\rf{Usowroll}. 

By defining two new scalar fields $\phi_1$ and $\phi_2$ as a linear combination  of the $\vp_1$ and $\vp_2$, such that we have a  transformation  orthogonal between these fields 
\be
\phi_1=\frac{A_1\a_2\vp_1+A_2\a_1\vp_2}{\sqrt{(A_1\a_2)^2+(A_2\a_1)^2}},\,\,\,\,\mbox{and}\,\,\,\,\,\phi_2=\frac{A_1\a_2\vp_1-A_2\a_1\vp_2}{\sqrt{(A_1\a_2)^2+(A_2\a_1)^2}}.
\ee
Note that from Eq.(\ref{FF1}) the new scalar field $\phi_1$ becomes an arbitrary constant, such that $\dot{\phi_1}=0$. Thus we can rewrite the new field $\phi_2$ in terms of the $\vp_2$ as
\be
\phi_2=\frac{C-2\Big(\frac{A_2\a_1}{A_1\a_2}\Big)\vp_2}{\sqrt{\Big(\frac{A_2\a_1}
{A_1\a_2}\Big)^2+1}}.
\lab{C}
\ee

In this way the effective potential can be write as a function of the new scalar field $\phi_2$ as

\be
U_{ eff}(\phi_2)= 
\left(\frac{f_1^2+2f_1g_1e^{\delta_0\phi_2+\delta_1}}{4\chi_2f_2(1+\epsilon f_1^2/f_2)}\right) \; ,
\lab{U-prime-minus1}
\ee
where the constants $\delta_0$ and  $\delta_1$ are given by 
$$
\delta_0=d_0d_1,\,\,\,\,\mbox{and}\,\,\,\delta_1=C[\a_1-d_0],\,\,\,\,\mbox{where}\,\,\,
d_0=\frac{A_1\a_2^2}{2A_2\a_1}\Big[\frac{A_2\a_1^2}{A_1\a_2^2}+1\Big],\,\,\,\,\,\mbox{and}\,\,\,\,\,d_1=\sqrt{\Big(\frac{A_2\a_1}{A_1\a_2}\Big)^2+1}\,.
$$
Notice that the shift $\phi_2 \rightarrow \phi_2 + \Delta$ has the same effect of just changing the value of $g_1$ according to 
$g_1 \rightarrow g_1 e^{\delta_0\Delta }$, so,  we would expect physical quantities not to depend on $g_1 $, except for its sign, if we restrict to inflation observables. Independence of
the absolute value of $g_1$ of physical quantities is equivalent to the independence of physical quantities under the shifts $\phi_2 \rightarrow \phi_2 + \Delta$,
that is that we do not care from where in the infinite plateau we start the slow roll inflation, invariance of physical quantities under  $\phi_2 \rightarrow \phi_2 + \Delta$  also means that the integration $C$ in \rf{C} should not affect any physical quantity relevant to the inflationary period.

Under slow roll approximation the motion equation for the new field becomes
\be
3\,H\,A_2\,\dot{\phi}_2\simeq-\frac{4\Big(\frac{A_2\a_1}{A_1\a_2}\Big)^2}{d_1^2}\,U_{eff,\phi_2}.\label{Fg1}
\ee
From Eqs.(\ref{SLR-def1}) and (\ref{Fg1}), we find that the slow roll parameter $\epsilon_H$ results
\be
\epsilon_H\simeq\frac{1}{A}\left(\frac{U_{eff,\phi_2}}{U_{eff}}\right)^2
\label{EE}
\ee
where the constant $A$
is defined as 
$$
A=\frac{A_2\,d_1^2}{4\left(\frac{A_2\a_1}{A_1\a_2}\right)^2}.
$$

Note that the two second slow roll parameters $\eta_1=\eta_2=\eta=-\ddot{\phi_2}/{(H\dot{\phi_2})}$ are  equivalent, since $\dot{\vp_1}\propto\dot{\vp_2}\propto{\dot{\phi_2}}$.

Therefore, we obtain that the slow roll parameter $\epsilon_H$
can  be written as
\be
\epsilon_H\simeq\frac{16g_1^2\delta_0^2 \,e^{2(\delta_0\,\phi_2+\delta_1)}}{A f_1^2}=\Big(\frac{16g_1^2\a_2^2}{A_2f_1^2}\Big)\left(\frac{A_2\a_1^2}{A_1\a_2^2}+1\right)^2\,\,e^{2(\delta_0\,\phi_2+\delta_1)}.
\ee
Here we have used that the effective potential $U_{eff}\simeq (f_1^2/f_2)/
(4\chi_2[1+\epsilon f_1^2/f_2])$. In order to find the scalar field at the end of the inflationary scenario, we consider that the slow roll parameter $\epsilon_H(\phi_2=\phi_{2 end})=1$ (or equivalently $\ddot{a}=0$) with which 
\be
e^{2\delta_0 \,\phi_{2 end}}=\Big(\frac{A_2f_1^2}{16g_1^2\a_2^2}\Big)\left(\frac{1}{\Big[\Big(\frac{A_2\a_1^2}{A_1\a_2^2}\Big)+1\Big]^2}\right)e^{-2\delta_1}.
\ee

Introducing the number of folds $N$ 
between two values of  times $t_{*}$ and $t_{end}$
(or analogously between two different values $\phi_{2*}$ and $\phi_{2 end}$) we have
\be
N = \int_{t_{*}}^{t_{\rm end}} H dt =
\int_{\phi_{2*}}^{\phi_{2 end}} \frac{H}{\dot{\phi}_2} d\phi_2 \simeq
- A\int_{\phi_{2*}}^{\phi_{2 end}} \frac{3H^2 }{U_{ eff,\phi_2}} d\phi_2 \simeq
-A \int_{\phi_{2*}}^{\phi_{2 end}} \frac{ U_{ eff}}{2 U_{ eff,\phi_2}} d\phi_2
\; ,
\lab{e-foldings}
\ee
and then we obtain that the number $N$ becomes
\be
N=\left(\frac{A_2f_1}{16g_1\a_2^2\Big[\Big(\frac{A_2\a_1^2}{A_1\a_2^2}\Big)+1\Big]^2}\right)\,\left[e^{-\delta_0\phi_{2end}-\delta_1}-e^{-\delta_0\phi_{2*}-\delta_1}\right].
\ee
Here we mention that the observational quantity $N$ should be evaluated when the cosmological
scale exits the horizon. In the following the subscript $*$ is utilized  to denote the epoch in which the cosmological
scale exits the horizon.

On the other hand, we will determine  the scalar and tensor
perturbations during the inflationary era for our model of a single field $\phi_2$ (since $\phi_1=$ cte). It is well known that for multifield inflation, the expressions for the scalar and tensor perturbations are modified in relation to  a single field, see Ref.\cite{Kaiser:2013sna}.  However from the relation  between the scalar fields given by Eq.(\ref{FF1}), the inflationary dynamics in our model is reduced to a single field $\phi_2$ and then  the scalar and tensor perturbations correspond to the standard formulas. Thus, from Ref.\cite{p1}
the power spectrum of the scalar perturbation ${\cP_S}$
under the slow-roll approximation for the new scalar field $\phi_2$ is given by
\begin{equation}
{\cP_S} = \left(\frac{H^2}{2\pi\,\dot{\phi_2}}\right)^2 \simeq\left(\frac{A^2}{96\pi^2}\,\frac{U_{eff}^3}{ U_{eff,\phi_2}^2}\right) ,\label{pec}
\end{equation}
and the observational  scalar spectral index $n_s$ is defined as
\be
n_s-1=\frac{d\ln \cP_S}{d\ln k}=
-6\epsilon_H+2\eta \; ,
\lab{ns}
\ee
where the slow roll parameters $\epsilon_H$ is defined by Eq.(\ref{EE}) and $\eta$ is given by $\eta\simeq (2/A)(U_{eff,\phi_2\phi_2}/U_{eff})$.

Also, it is well known that the generation of tensor perturbations during the epoch of 
inflation  would generate gravitational waves. In this sense, the spectrum of the tensor perturbations
$\cP_T$ is given by\cite{p1} 
\begin{equation}
\cP_T=\left(\frac{H}{\pi}\right)^2 \simeq \frac{U_{eff}}{6\pi^2}\; ,
\label{Pt}
\end{equation}
and its  tensor spectral index $n_T$ can be described  in terms of
the slow parameter $\epsilon_H$ as $ n_T=\frac{d\ln \cP_T}{d\ln k}=-2\epsilon_H$.  Besides, 
an important observational quantity corresponds to the tensor-to-scalar ratio
$r=\frac{\cP_T}{\cP_S}$. Here we emphasize  that these quantities should be evaluated during the epoch in which the cosmological scale exits the horizon and in particular at $\phi_2=\phi_{2*}$.

In this way, from Eq.(\ref{pec}) the power spectrum of the scalar perturbation ${\cP_S}$
under the slow-roll approximation in terms of the number of $e-$folds yields
\be
{\cP_S}(N)\simeq k_0\,\left[F_1-\frac{N}{F_2}\right]^2,\,\,\,\,\mbox{where}\,\,\,\,\,\,k_0=\frac{A^2 f_1^4}{96\times(4\pi)^2\chi_2f_2g_1^2\delta_0^2(1+\epsilon f_1^2/f_2)},
\ee
$$
F_1=\frac{4g_1\a_2}{f_1\sqrt{A_2}}\left[\frac{A_2\a_1^2}{A_1\a_2^2}+1\right],\,\,\,\,\,\mbox{and}\,\,\,\,\,F_2=\left(\frac{A_2f_1}{16g_1\a_2^2\Big[\Big(\frac{A_2\a_1^2}{A_1\a_2^2}\Big)+1\Big]^2}\right),
$$
respectively.

Also, the scalar spectral index $n_s$ in terms of the number of $e-$folds $N$ becomes
\be
n_s=1-\frac{8g_1\delta_0^2}{A f_1}\,\left(F_1-\frac{N}{F_2}\right)^{-1}\,\left[3g_1\left(F_1-\frac{N}{F_2}\right)^{-1}+1\right].
\lab{ns}
\ee
From this scalar spectral index $n_s$ we find that the  solution for the parameter $\a_2$  is given by
\be
\a_2=\frac{\sqrt{A_2}(1+4(n_s-1)N)}{16NL_1}+\frac{1}{16}\left[+\sqrt{L_2}+\sqrt{L_3}\right],
\label{a2a}
\ee
where the values of $L_1$, $L_2$ and $L_3$ are defined as 
$$
L_1=[1+2(n_s-1)N],\,\,\,\,\,\,\,\,\,L_2=\frac{A_2(1-12f_1N-24f_1(n_s-1)N)}{N^2L_1^2},
$$
and
$$
L_3=\frac{-256\a_1^2A_2N^2L_1^2+2A_1A_2N[1-2(2+3f_1-2n_s)L_1]-2A_1\sqrt{A_2}N[L_1(3+4(n_s-1)N)]\,L_4}{A_1N^2L_1^2},
$$
with $L_4$ given by
$$
L_4=\sqrt{\frac{A_2(1-12f_1N-24f_1(n_s-1)N^2)}{N^2L_1^2}}.
$$

We mention that in the solution  for the parameter $\a_2$ given by Eq.(\ref{a2a}), we have considered the solution with the positive signs.

By  using that the tensor to scalar ratio $r$ under the slow roll approximation is given by $r\simeq16\epsilon_H$ we find
\be
r\simeq\frac{64g_1^2\delta_0^2}{A f_1^2}\,\left[F_1-\frac{N}{F_2}\right]^{-2}.
\lab{r}
\ee

From this expression for the tensor to scalar ratio, we find that the parameter $\alpha_1$ becomes
\be
\a_1=\frac{1}{2}\sqrt{\frac{A_1\,\a_2}{\sqrt{A_2}}\left[\frac{2}{N\sqrt{r}}+\frac{1}{N}-\frac{4\a_2}{\sqrt{A_2}}\right]},
\ee
where we have considered the positive sign for the solution of $\a_1$.  Also, as we anticipated, physical quantities cannot depend on the magnitude of $g_1$ and in fact we  note that this result obtained  for the parameter $\a_1$ does not depend of the parameter $g_1$ analogously as occurred for the $\a_2$ and neither $n_s$ as given by \rf{ns}, $r$  as given by  \rf{r}, etc.

In particular assuming the values $f_1=f_2=10^{-7}$, $\epsilon=10^{-4}$ and $\chi_2=1$, together with ${\cP_S}\simeq 10^{-9}$, $n_s=0.967$ and $r=0.039$ at $N=60$, then numerically we find that the parameters $\a_1$, $\a_2$ and $b_1$ result  $\a_1\simeq 10^{-9}$, $\a_2\simeq 10^{-16}$ and $b_1\simeq -1$, respectively. Here we note that the special value of the parameter $b_1 \simeq -1$ we have found from inflation  is also in the allowed values found for a consistent stable  emergent universe condition, see Fig.\ref{fig1}.

In order to find the consistency relation i.e., the relation between the tensor to scalar ratio $r$ and the scalar spectral index $n_s$, we can combine Eqs.\rf{ns} and \rf{r} obtaining 
\be
n_s(r)=n_s=1-\delta_0\sqrt{\frac{r}{A}}\left[\frac{3f_1\sqrt{A\,r}}{8\delta_0}+1\right].
\ee

\begin{figure}
 	\centering
\includegraphics[width=0.68\textwidth]{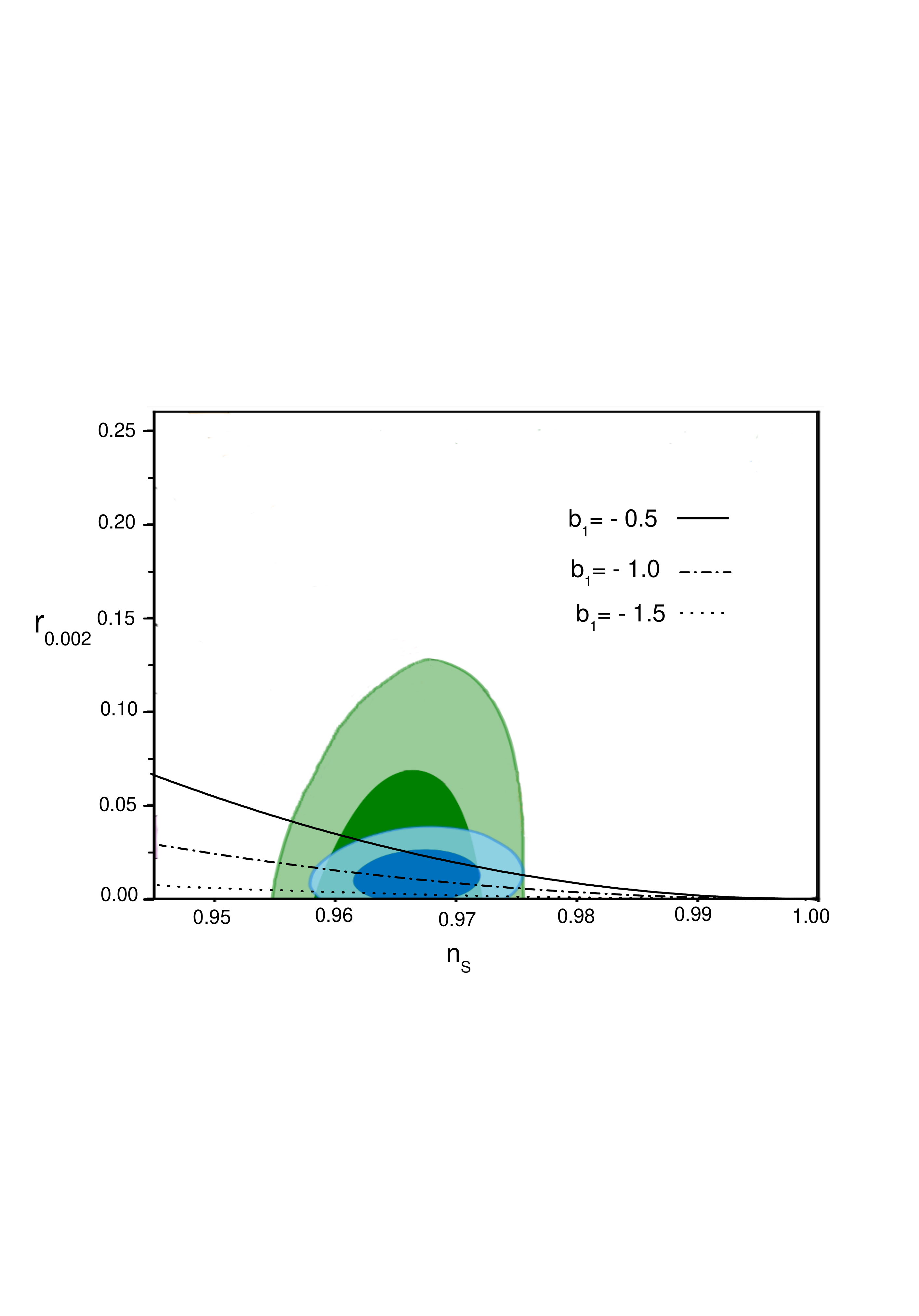}
{\vspace{-3.5cm}
\caption{\it{ {The plot shows the tensor to scalar ratio $r$ versus the scalar spectral index $n_s$ (consistency relation $r=r(n_s)$) for  three values of the dimensionless  parameter $b_1$. For $b_1=-0.5$ solid line, $b_1=-1$ dash-dot line and $b_1=-1.5$ dots line, respectively. 
 Here we have fixed the values $\chi_2=1$, $\epsilon=10^{-4}$ (in units of $M_{Pl}^{-2}$),  $f_1=f_2=10^{-8}$ (in units of $M_{Pl}^{4}$), $\alpha_1=4\times10^{-9}$  and $\alpha_2=10^{-16}$ (in units of $M_{Pl}^{-1}$). Here we have used the two-dimensional marginalized joint confidence contours for $(n_s,r)$ at 68\% and 95\% levels of confidence from the latest BICEP/Keck results \cite{BICEP:2021xfz}. \label{fig4}} }}
}	
\end{figure}

In figure \ref{fig4} we show the dependence of the tensor to scalar ratio on the scalar spectral index for different values of the parameter $b_1$ studied during the emergent epoch (see Fig.\ref{fig1}) from the latest BICEP/Kerck data \cite{BICEP:2021xfz}. From the Ref.\cite{BICEP:2021xfz}, two-dimensional marginalized constraints on inflationary parameters $r$ and the scalar spectral index $n_s$ defined at $k_0=0.002$Mpc$^{-1}$. The latest BICEP/Keck results places stronger limits on the tensor to scalar ratio shown in blue (at 68\%  blue region and 95\%  light blue region levels of confidence) and the green color corresponds to 
 the two-dimensional marginalized constraints  obtained in Ref.\cite{Planck:2018vyg}.  We noted from Fig.(\ref{fig4}) that the parameter $b_1$, which lies in the range $-2<b_1<0$ in order to satisfy the stable emergent universe solution, it  is well supported by the the latest BICEP/Keck  data.

\section{First Dark region with consistent generation of DE, DM and stiff matter}\label{DEDM}

In this section we will analyze the second flat region in which the scalar field $\varphi_2\rightarrow -\infty$  and it corresponds to the dark energy sector region and the effective potential in this region is given by Eq.\rf{U-minus2}. In order to study this scenario that includes  dark matter together with dark energy, we use the $k-$essence model in which dark matter appears naturally as was shown in Ref.\cite{Scherrer:2004au}, see also Ref.\cite{guendelmanKineticDEDM} for other examples of this effect in TMT.

In this context, the equation of motion for the scalar field $\varphi_1$ is reduced to
$$
\frac{d}{dt}\left[a^3\dot{\varphi}_1(A_{1g}+2B_{1g}X_1+B_{12g}X_2)\right]=0.
$$
The obvious solution corresponds to $\dot{\varphi}_1=0$, together with  the term $(A_{1g}+2B_{1g}X_1+B_{12g}X_2)\neq0$. In this way we can consider that the term $X_1=0$. Thus,  under a small perturbation we can assume  that
$$
X_1=0+\delta_1,\,\,\,\,\,\,\mbox{then}\,\,\,\,\,\,\,\frac{d}{dt}(a^3\delta_1^{1/2})=0,
$$
and the solution for $X_1$ can be written as
\be
X_1=\delta_{01}\left(\frac{a}{a_0}\right)^{-6},\label{X1b}
\ee
where $\delta_{01}$ corresponds to a positive constant.

On the other hand, the equation of motion for the scalar field $\varphi_2$ is given by 
\be
\frac{d}{dt}\left[a^3\dot{\varphi}_2(A_{2g}+2B_{2g}X_2+B_{12g}X_1)\right]=0.
\label{A2g}
\ee
In this situation we do not consider the solution $\dot{\varphi}_2=0$, instead we assume  the case where  $(A_{2g}+2B_{2g}X_2+B_{12g}X_1)=(A_{2g}+2B_{2g}X_2)=0$, where we have used that  $X_1=0$. Thus the obvious solution for $\dot{\varphi}_2$ can be written as
\be
X_2=-\frac{A_{2g}}{2B_{2g}}=X_{20}=\mbox{positive constant}.\label{X2}
\ee
In this case we must have $b_2^2>4g_2\epsilon(1+b_2g_1/g_2)$ with which the term $B_{2g}$ is negative and $A_{2g}>0$ (or $-2<b_2g_1/g_2$).

Following Ref.\cite{Scherrer:2004au} we can show  that the solution given by Eq.(\ref{X2}) is stable under a small perturbation. In this form, we can perturb this solution as 
\be
X_2=-\frac{A_{2g}}{2B_{2g}}+\delta_2=X_{20}+\delta_2,\label{Del}
\ee
with $\delta_2\ll X_{20}$. In this way, replacing Eq.(\ref{Del}) into Eq.(\ref{A2g}) and expanding to order one in $\delta_2$ yields
$$
\frac{d}{dt}\,(a^3\,\delta_2)=0, \,\,\,\,\mbox{then}\,\,\,\,\,\delta_2\propto\,a^{-3}.
$$
Thus, the perturbative  solution given by Eq.(\ref{Del}) results
\be
X_2(a)=X_{20}\left[1+\delta_{02}(a/a_0)^{-3}\right], \label{X2a}
\ee
where  $\delta_{02}$ and  $a_0$ are two new constants. Besides, as $\delta_2\ll X_{20} $, then we have $\delta_{02}(a/a_0)^{-3}\ll 1$.

Replacing Eqs.(\ref{X1b}) and (\ref{X2a})   into the energy density $\rho$ we have
\be
\rho(a)\simeq\rho_0+\rho_1\,\left(\frac{a}{a_0}\right)^{-3}+\rho_2\,\left(\frac{a}{a_0}\right)^{-6},\label{RRR}
\ee
where $\rho_0$ corresponds to the energy density associated to dark energy and it is given by
$$
\rho_0=\frac{1}{4}\frac{A_{2g}^2}{B_{2g}}+ U_{\rm eff\,g}=-\frac{1}{\chi_2}\left[\frac{b_2g_1+g_2}{b_2^2-4\epsilon(b_2g_1+g_2)}\right]>0.
$$
Note that in Eq.(\ref{RRR}) we have neglected  the term $(a/a_0)^{-12}$ since during the evolution of present universe this term is highly suppressed.

Also, we note that as the term  $b_2^2>4\epsilon(b_2g_1+g_2)$ (see Eq.(\ref{X2})) then  we have that $b_2g_1+g_2<0$, since $\epsilon>0$ and  from Eq.\rf{A-B-minus2} we have that the term $A_{2g}>0$, then we get that $b_2g_1/2+g_2>0$. In this form, assuming that the parameter $g_2$ is positive, we find that the constraint for the ratio $b_2g_1/g_2$ becomes
\be
-1>\frac{b_2g_1}{g_2}>-2.
\ee
This result suggest that the parameter $b_2$ is a  positive quantity, since during the inflationary epoch we consider that the effective potential does not have bump  when $g_1<0$.

Now the second term of Eq.(\ref{X2a}) represents to the dark matter and the quantity $\rho_1$ is defined as
$$
\rho_1=\frac{\delta_{02}\,A_{2g}^2}{B_{2g}}.
$$
We note that as $B_{2g}$ is a negative quantity, then we impose that the constant $\delta_{02}<0$, in order to obtain that the dark matter to be positive and also $\vert\delta_{02}\vert\ll 1$. 

The last term corresponds to stiff equation of state where the quantity $\rho_2$ is given by
$$
\rho_2=\,\frac{3\,A_{2g}^2}{4\,B_{2g}}\,\,\delta_{02}^2+A_{1g}\delta_{01}-\frac{3B_{12g}\delta_{01}A_{2g}}{2B_{2g}},
$$
and this parameter corresponds to a positive quantity.

Also, we find that the pressure in terms of the scale factor results
\be
p(a)\simeq p_0+p_1\left(\frac{a}{a_0}\right)^{-3}+p_2\left(\frac{a}{a_0}\right)^{-6}\simeq p_0+p_2\left(\frac{a}{a_0}\right)^{-6}\label{p12},
\ee
where the quantities $p_0$, $p_1$ and $p_2$ are defined as  
$$
p_0=-\frac{A_{2g}^2}{4B_{2g}}-U_{\rm eff\,g}=-\rho_0, \,\,\,\,\,\,\,\,\,p_1=X_{20}\delta_{20}[A_{2g}+2B_{2g}X_{20}]=0,\,\,\,\,
$$
and 
$$
p_2=\delta_{01}\left[A_{1g}-\frac{B_{12g} A_{2g}}{2B_{2g}}\right]+\frac{A_{2g}^2\delta_{02}^2}{4B_{2g}}.
$$
As before in Eq.(\ref{p12}) we have neglected the term $(a/a_0)^{-12}$.
Also, in order to obtain an adequate  stiff equation of state, which is necessary from  energy momentum conservation (and  therefore for the consistency of the equations of motion),  in which the coefficients $\rho_2=p_2$. In this form,  we find that the relation between the constants $\delta_{01}$ and $\delta_{02}$ from the condition given by  stiff equation of state yields
\be
\delta_{01}=\left(\frac{ A_{2g}}{2B_{12 g}}\right)\,\delta_{02}^2,
\lab{consistency}
\ee
and then we have that $\rho_2=p_2=\delta_{01}A_{1g}$.
Note that in order to obtain $\delta_{01}>0$, we have that from Eq.(\ref{B12}) the coefficient  
\be
B_{12g}=\frac{g_1^2+8\epsilon\chi_2^2[g_1(b_2-\epsilon g_1)+g_2]}{4\chi_2(\epsilon g_1^2+g_2)}>0.\label{B12ga}
\ee
We see first that for $\epsilon =0$ the above Eq.(\ref{B12ga}) is of course satisfied. Now 
 taking $\epsilon >0$, when we start to increase $\epsilon $ towards bigger positive values, we will reach a critical value of $\epsilon$ beyond which the condition breaks down i.e., when  $B_{12g}=0$. Thus,  we have then that the above condition holds for  $0< \epsilon < \epsilon_c$, where the critical value of $\epsilon$ becomes
$$\epsilon_c = \frac{2g_2(\frac{b_2g_1}{g_2}+1)\chi_2^2 \pm\sqrt{2g_1^4\chi_2^2 + 4(b_2g_1 + g_2)^2 \chi_2^4}}{4 g_1^2\chi_2^2}.$$
Since $-1>\frac{b_2g_1}{g_2}>-2$, the first term in the numerator of $\epsilon_c$ is negative and then  we must take the plus sign in the second positive term in the above equation so that $\epsilon >0$ is satisfied.

In conclusion, we can obtain definite   ranges in the  parameter-space for  a consistent generation of DE, DM and stiff component from multi fields dynamics.

On the other hand, the equation of state (EoS) or EoS parameter $w=p/\rho$ in our model as a function of the scale factor becomes
\be
w(a)=\frac{-\Omega_0+(1-\Omega_0-\Omega_1)(a/a_0)^{-6}}{\Omega_0+\Omega_1(a/a_0)^{-3}+(1-\Omega_0-\Omega_1)(a/a_0)^{-6}},\label{ww}
\ee
where the parameters $\Omega_0$ and $\Omega_1$ associated to the dark energy and dark matter at the present time are defined as
$$
\Omega_0=\frac{\rho_0}{6H_0^2},\,\,\,\,\mbox{and}\,\,\,\,\,\,\Omega_1=\frac{\rho_1}{6H_0^2},
$$
with $H_0$ the Hubble parameter at the present epoch. Here we have used the Friedmann equation $6H^2=\rho$, in which $1-\Omega_0-\Omega_1=\Omega_2=\rho_2/6H_0^2$ at the present era. From Eq.(\ref{ww}) we note  that in the far future  in which the scale factor is large, the EoS parameter $w\rightarrow -1$, i.e., the universe is dominated by a cosmological constant.

\begin{figure}
 	\centering
\includegraphics[width=0.49\textwidth]{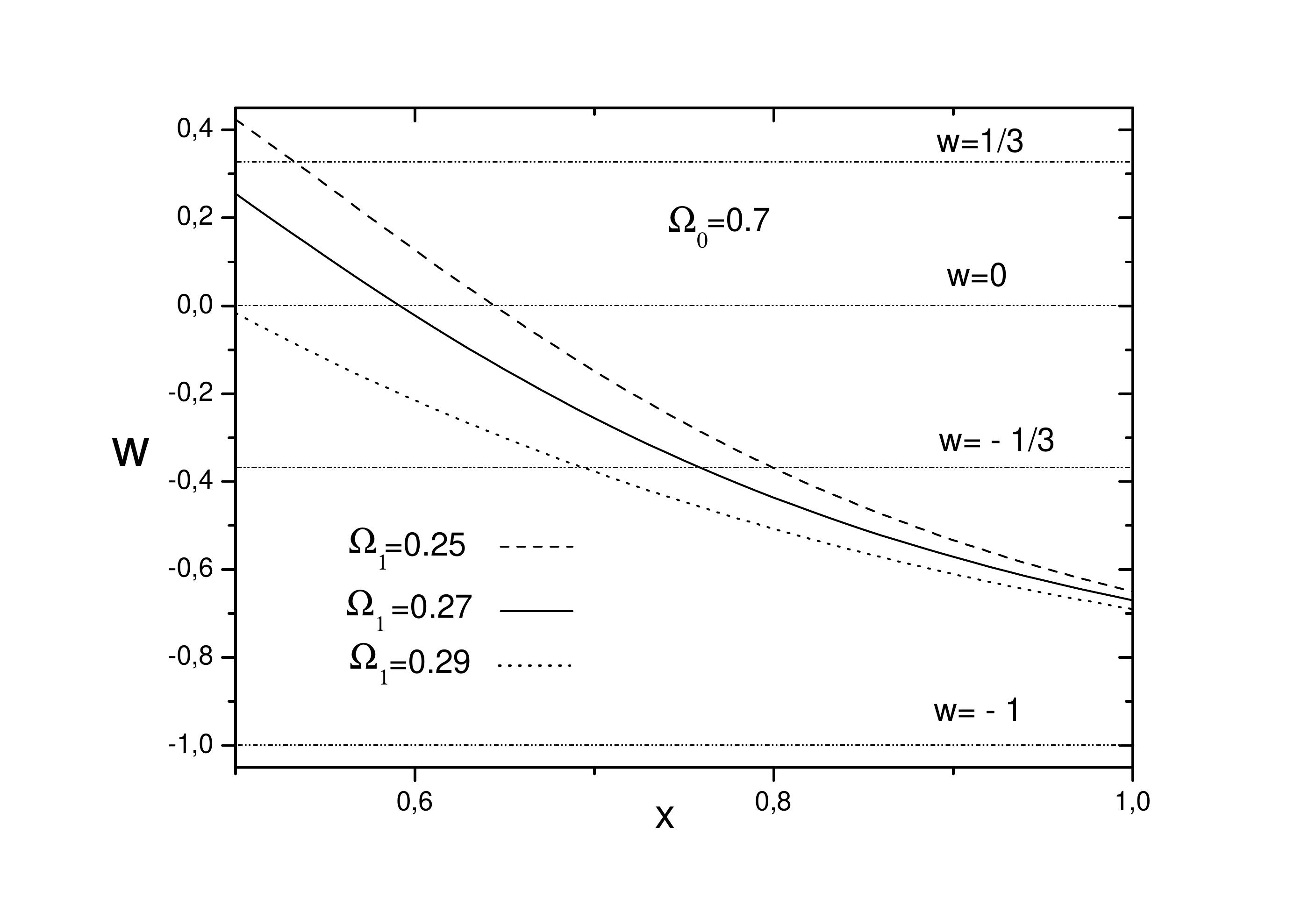}
\includegraphics[width=0.49\textwidth]{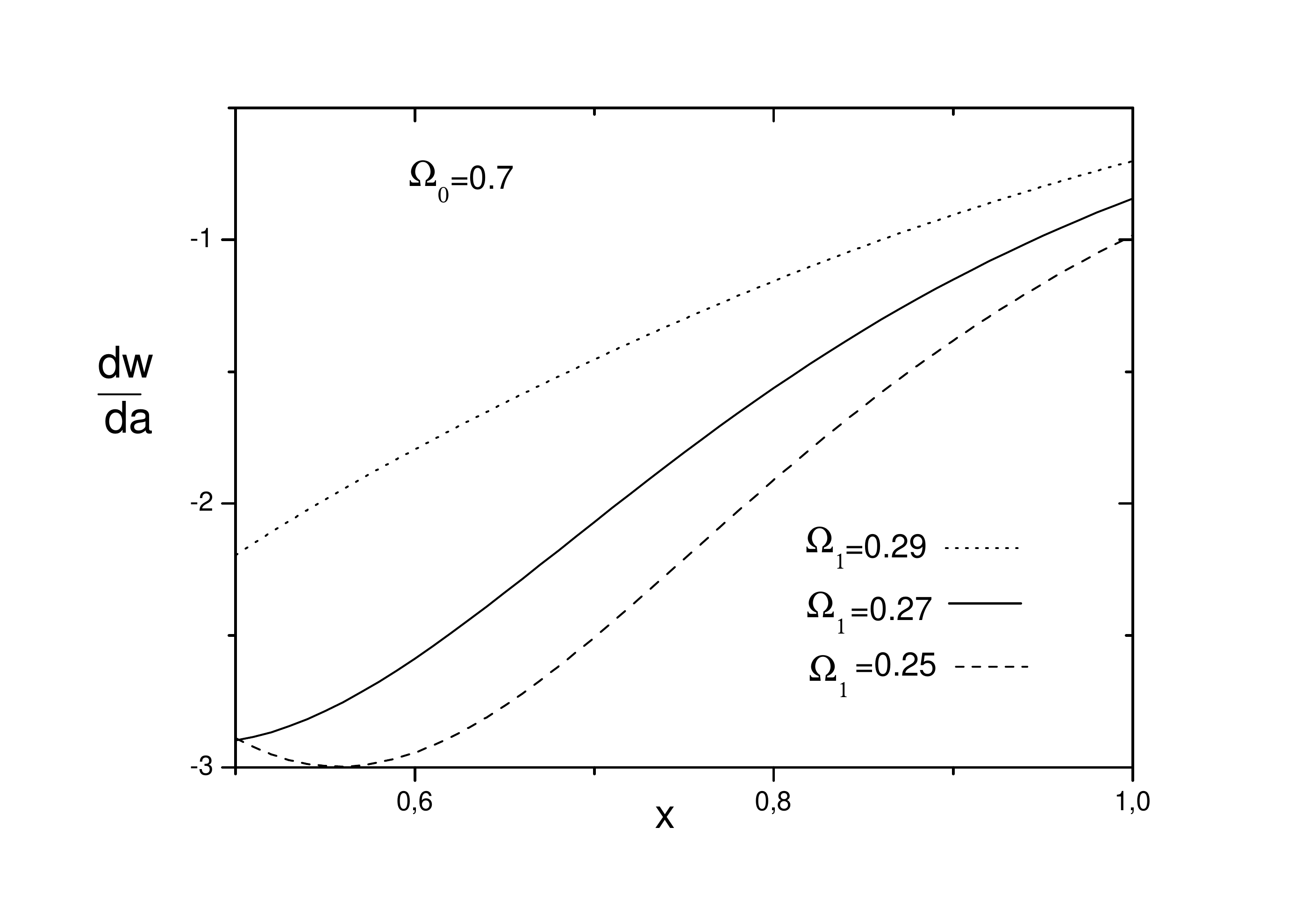}
\caption{\it{ {The left panel shows the evolution of the EoS parameter as a function of the normalized   scale factor $x=a/a_0$. The right panel shows the behaviour of $dw/da$ versus $x$. Here we have considered    different values of the density parameter associated to the dark matter $\Omega_1$ and also we have fixed  the density parameter of the dark energy to $\Omega_0=0.7$.} }}

 	\label{fig2}
\end{figure}

In Fig.\ref{fig2} (left panel) we show the evolution of the EoS parameter $w(x)$ versus the normalized  scale factor $x=a/a_0$ defined by Eq.(\ref{ww}) for same values of the parameter $\Omega_1$. Here we have fixed that the density parameter associated to dark energy $\Omega_0=0.7$. Additionally, we have added different horizontal lines associated to the constant values of the EoS parameter $w=1/3, w=0, w=-1/3$ and $w=-1$, respectively. This help us distinguish the radiation dominated era, the matter domination era and the dark energy domination era in terms of the parameter $\Omega_1$. In this sense, from Fig.\ref{fig2} (left panel) we observe that the end of the radiation dominated, matter dominated universe and dark energy occurs later for the situation in which the parameter associated to dark matter $\Omega_1$ decrease. Thus, for  example in the case $\Omega_1=0.25$  the EoS $w=0$ (matter domination) takes place for a scale factor $x\simeq 0.67$ while  for the value $\Omega_1=0.29$ occurs before  at $x\simeq 0.5$.  
From the plot we note that when we decrease the density parameter associated to the dark matter $\Omega_1$ the EoS parameter increase. Also, we observe that for $\Omega_1<0.29 $ (more stiff matter) and for values of  the scale factor $x<0.7$, the universe does not present an accelerated  phase, since the EoS parameter becomes positive. Thus,  we note that when we have a very small quantity of stiff matter ($\Omega_1\ge 0.29$) the universe always  presents an accelerated scenario in the plot region for the scale factor. In order to account of the transition from deceleration expansion to acceleration expansion, we have that from Eq.(\ref{ww}), the situation  in which the pressure can be considered negligible or matter domination ($\omega\sim 0$) occurs when the scale factor $a/a_0=x\geqslant (\Omega_0/(1-\Omega_0-\Omega_1))^{1/6} $ up to the value $a/a_0=x<(\frac{\Omega_1[1+\sqrt{1+32\Omega_0(1-\Omega_0-\Omega_1)/\Omega_1^2}]}{4\Omega_0})^{1/3}$ where $w>-1/3$, see Fig.\ref{fig2}. During this period the energy density given by Eq.(\ref{RRR}) can be approximated to $\rho\propto a^{-3} $ and the scale factor evolve as $a\propto t^{1/2}$. However, for values of the scale factor $a/a_0=x\gtrsim(\frac{\Omega_1[1+\sqrt{1+32\Omega_0(1-\Omega_0-\Omega_1)/\Omega_1^2}]}{4\Omega_0})^{1/3}$, the universe presents an accelerated scenario ($\ddot{a}>0$) in which the energy density can be considered as a constant. In this sense, from the  EoS parameter found in our model (Eq.(\ref{ww})) we can account of the matter dark energy domination  transition may precede the deceleration to acceleration transition.  The study of this interrelation of the epochs of deceleration-acceleration transition from  explicit EoS parameters was analyzed in detail  \cite{Nojiri:2006jy} and references therein.

In the right panel of Fig.\ref{fig2} we shows the evolution of the derivative $dw/da$ versus the normalized scale factor $x$. From this plot we observe that the model corresponds to a tracking freezing model where the derivative $dw/da$ is negative. Also we note that when we decrease the values of the density parameter related to dark matter the quantity $dw/da$ results more negative.

It is interesting to study the behaviour of the EoS parameter $w(a)$ and in particular the derivative  $dw/da$ related  to   the dark epoch into the second flat region. In this way from Eq.(\ref{ww}) we find that the quantity $dw/da$ becomes
\be
\frac{dw}{da}=-3x^2\left(\frac{(1-\Omega_0-\Omega_1)(\Omega_1+4\Omega_0x^3)+\Omega_0\Omega_1 x^6}{[1+\Omega_1(x^3-1)+\Omega_0(x^6-1)]^2}\right)<0,\,\,\,\,\,\mbox{with}\,\,\,\,x=\frac{a}{a_0}.
\ee
This result shows that the behaviour of our model corresponds to a tracking freezing model since $dw/da<0$, see Refs.\cite{Ar1,Ar2}. 

On the other hand, from Friedmann equation
$6H^2=\rho$, we can obtain that the scale factor as a function of the cosmological time $t$ i.e., $a(t)$ results \cite{Chavanis:2014lra}
\be
\frac{a(t)}{a_0}=\left[\left(\frac{\Omega_1}{\Omega_0}+2\sqrt{\frac{(1-\Omega_0-\Omega_1)}{\Omega_0}}\right)\sinh^2\left[\frac{3\sqrt{\Omega_0}(H_0t+C)}{2}\right]+\sqrt{\frac{(1-\Omega_0-\Omega_1)}{\Omega_0}}\left(1-e^{-3\Omega_0^{1/2}(H_0t+C)}\right)\right]^{1/3},\label{a12}
\ee
where $C$ corresponds to an integration constant. Note that in the absence of stiff matter ($1-\Omega_0-\Omega_1=\Omega_2=0$) the scale factor $a(t)\propto \sinh^{2/3}(3\sqrt{\Omega_0}(H_0t+C)/2)$, see Ref.\cite{Frieman:2008sn}

Also, we can find the scalar fields in terms of the scale factor. Thus, for the scalar field $\varphi_1$ we have that
from Eq.(\ref{X1b}) that the speed of the scalar field $\varphi_1$ can be written  as
$$
\dot{\varphi}_1=Hx\frac{d\varphi_1}{dx}=\pm\frac{\sqrt{2\delta_{01}}}{x^3},
$$
 and its solution yields
\be
\varphi_1(x)=C_1\mp\frac{\sqrt{2\delta_{01}}}{3H_0\sqrt{1-\Omega_0-\Omega_1}}\mbox{Arctanh}\left[\frac{\Omega_1x^3+2(1-\Omega_0-\Omega_1)}{2\sqrt{1-\Omega_0-\Omega_1}\sqrt{x^3(\Omega_1+\Omega_0\,x^3)+(1-\Omega_0-\Omega_1)}}\right].\label{a13}
\ee

Analogously for the scalar field $\varphi_2$, we find that the solution as a function of the normalized  scale factor $x=a/a_0$ can be expressed as
\be
\varphi_{2}(x)=C_1\pm\frac{2\sqrt{2X_{02}}}{3H_0}\sqrt{\frac{\bar{\Omega}-\Omega_1-2\Omega_0x^3}{\bar{\Omega}}}\left(\frac{g_1(x)-g_2(x)}{g_3(x)}\right),\label{a14}
\ee
where $C_1$ is another integration constant and the functions $g_1(x)$, $g_2(x)$ and $g_3(x)$ are defined as
$$
g_1(x)=\sqrt{\frac{\Omega_0(x^3+\delta_{02})}{2\delta_{02}-\Omega_1-\bar{\Omega}}}
\,(\bar{\Omega}^2+\Omega_1\bar{\Omega}+\Omega_0\Omega_1x^3+\Omega_0\bar{\Omega}x^3)\,{\cF}[\bar{x},m_1],
$$
$$
g_2(x)=\delta_{02}\Omega_0(\Omega_1+\bar{\Omega}-2\delta_{02}\Omega_0)\,(\Omega_1+\bar{\Omega}+2\Omega_0x^3)\,\sqrt{\frac{1}{\Omega_1+\bar{\Omega}-2\Omega_0\delta_{02}}}\,\sqrt{\frac{(\delta_{02}+x^3)\Omega_0}{2\Omega_0\delta_{02}-\Omega_1-\bar{\Omega}}}\,\Pi[m_2,\bar{x},m_1],
$$
and
$$
g_3(x)=\Omega_0(\Omega_1+\bar{\Omega})\,\sqrt{\delta_{02}+x^3}\,\sqrt{\frac{\Omega_1+\bar{\Omega}+2\delta_{02}\Omega_0}{\Omega_1+\bar{\Omega}-2\delta_{02}\Omega_0}}\sqrt{(1-\Omega_0-\Omega_1)+\Omega_1x^3+\Omega_0x^6}.
$$
Here the functions $\cF$ and $\Pi$ correspond to the Elliptic functions of the first and second kind, respectively, see \cite{Libro}. Also, the quantities $\bar{x}$, $m_1$, $m_2$,  and $\bar{\Omega}$ are given by
$$
\bar{x}=\mbox{Arcsin}\left[\sqrt{\frac{\Omega_1+\bar{\Omega}+2\Omega_0x^3}{\Omega_1+\bar{\Omega}-2\delta_{02}\Omega_0}}\right],\,\,\,\,\,\,
m_1=\frac{\Omega_1+\bar{\Omega}-2\delta_{02}\Omega_0}{2\bar{\Omega}},\,\,\,m_2=\frac{\Omega_1+\bar{\Omega}-2\delta_{02}\Omega_0}{\Omega_1+\bar{\Omega}},
$$
and $\bar{\Omega}=\sqrt{\Omega_1^2-4\Omega_0(1-\Omega_0-\Omega_1)}$.

We mention that in order to obtain the scalar fields as a function of the cosmological time, we need to replace the normalized scale factor given by  Eq.(\ref{a12}) into Eqs.(\ref{a13}) and (\ref{a14}) to find $\varphi_1(t)$ and $\varphi_2(t)$, respectively.

\section{Summary and Discussion of Possible future lines of research}\label{discuss}

In summary, in this paper we have explored a new approach to study all the most relevant aspects of the evolution of the universe using a scale invariant theory with metric independent measures which employs two scalar fields. In this sense, we have analyzed in the early universe the scenarios of the emergent and inflationary epochs. Also we have studied the dark sector considering the DE, DM and stiff matter during the late universe. In relation to
the scale invariance is spontaneously broken by the integration of the equations of motion and going to the Einstein frame, we find a theory with an effective potential for the scalar fields that displays three flat regions and corresponding K-essence terms.
We use each of theses flat regions to describe the different phases of the universe. The higher energy density describes the very early universe, first a non singular emergent universe which is then followed by an inflationary era. 

For the emergent universe we have analyzed the stability of the emergent solution from the perturbative analysis. In this context because of the non triviality   of the equation associated to the  parameter $\omega^2$ we have considered a numerical treatment. In particular for different values of the parameter-space,  we have found that the parameter $b_1$ is negative and its value is in the range $-2<b_1<0$, see Fig.\ref{fig1}. Here we have  obtained  that this range for the parameter $b_1$
does not depend of the values of $f_1$ and $f_2$.

In the context of the inflationary epoch, we have considered the slow roll approximation. Under this approximation we have defined two new scalar fields $\phi_1$ and $\phi_2$ from  a transformation orthogonal in which $\dot{\phi}_1=0$ and then the effective potential reduces to a single scalar field $\phi_2$, see Eq.\rf{UUU}. Also, we have noted that 
the effective potential \rf{UUU}
interpolates between the vacuums with asymptotic values $\frac{f_1^2}{4\chi_2(f_2+\epsilon f_1^2)}$ and $\frac{g_1^2}{4\chi_2(g_2+\epsilon g_1^2)}$ independent of the choice of the relative signs of $g_1$, and $f_1$. However, we have chosen the sign of  $f_1g_1$ negative, so that the potential smoothly rolls down from one vacuum to the other without an intermediate bump. From the observational parameters; scalar power spectrum, scalar spectral index and the upper bound of the ratio tensor to scalar,  we have found numerically the values of the parameters $\alpha_1$, $\alpha_2$ and $b_1$. In this context, we have obtained that the constraint imposes for the inflationary epoch for the parameter $b_1$ is in the range found  from the stability analysis given for the  emergent universe.

 The second flat region describes 
 a dark sector with  DE and DM and a quickly decaying stiff component as well. The DE,  DM and stiff components owe its existence to the K-essence induced by multi measure theory. It is interesting that a consistency condition for the generation of the DE, DM and stiff components which correlates the perturbations of the two scalar fields with respect to a certain background solution  is obtained
in \rf{consistency} from the condition found  by stiff equation of state $\rho_2=p_2$. From the evolution of the EoS parameter $w$ versus the normalized scale factor $x=a/a_0$, we have noted that when we decrease the density parameter associated to dark matter  $\Omega_1$, the EoS parameter increase for the same value of $x$, see Fig.\ref{fig2} (left panel). Also, we have found that the derivative $dw/da$ is negative with which this model corresponds to a tracking freezing model, see Fig.\ref{fig2} (right panel). Besides, we have determined the solution for the scale factor as a function of the cosmological time together with the solutions  for the scalar fields in term of the normalized scale factor.

There is also a third flat region,  which could represent a future DE  sector, or may be represents the present state of the universe.
Among the  issues  that should be studied further, we mention first  whether the universe makes use of the third flat region with asymptotic  behavior \rf{U-M}, \rf{A-B-M} and \rf{B12-M}. In this case, we should connect the first dark region described in section VI with the one satisfying the asymptotic behavior \rf{U-M}, \rf{A-B-M} and \rf{B12-M}, in order to study the early dark epoch 
 and its transition to a late DE era.

For this purpose we can construct a simplified effective potential that interpolates between these two vacuums, that could be simply achieved by considering the effective potential given by  Eq.\rf{U-eff} and ignoring the $f$-terms, such that 
\be
U_{\rm eff}  =  \frac{(g_1 e^{-\a_2\vp_2} -M_1)^2}{4\chi_2\,\Bigl\lb
g_2 e^{-2\a_2\vp_2} + M_2 + \eps (g_1 e^{-\a_2\vp_2}-M_1)^2\Bigr\rb} \; .
\lab{U-gMeff}
\ee
By considering the analysis described  in the inflationary epoch (section V), we know that the potential \rf{U-gMeff} may connect the two vacuums  through a slow roll process, and will also contain a minimum at zero if $(g_1 e^{-\a_2\vp_2} -M_1)^2 = 0$ at some point, which would require  $M_1<0 $ , since we have determined that  $g_1<0 $. However,  we will assume that the parameter  $M_1>0$ in which case the potential will contain a bump and the only way to connect these two vacuums  will be through tunnelling, which reminds us of the original (or old) inflationary universe by Guth \cite{Guth}. This scenario, where two DE states are connected through bubble nucleation was considered in \cite{sloth}, \cite{sloth2} to formulate a possible resolution of the $H_0$ problem and we could consider it also in a future research for our model.

In this case one may ask what may happen to the dark matter generated in the first DE  region, could the DM remain as islands or impurities of the previous  vacuum?, or may be this  DM collapses into black holes before the transition to the final vacuum takes place, and then of course the black holes are resilient and should remain even after the transition to the final vacuum state. These issues could be studied in a future research.

The second subject concerning  we want to discuss concerning possible future lines of research is a very simple possible variation of the model that could lead to interesting results in the unification of the early and current universe. As we have seen in the previous sections, the best results are obtained for $\epsilon$ small, so it is worthwhile to study $\epsilon= 0$ . Furthermore, it is quite interesting that for  
 $\epsilon= 0$, we can also include a term, that which when considered in Einstein frame, produces just a shift in the effective potential, that is it just produces a the same effect of a cosmological term in Einstein frame.

 The slightly modified theory, where the $\epsilon$ is eliminated and a $\Lambda_0$ term is added and then the new action can be written as
 \be
S = \int d^4 x\,\P_1 (A) \Bigl\lb R -2 \Lambda_0 \frac{\P_1 (A)}{\sqrt{-g}}  + L^{(1)} \Bigr\rb +  
\int d^4 x\,\P_2 (B) \Bigl\lb L^{(2)}  + 
\frac{\P (H)}{\sqrt{-g}}\Bigr\rb \; ,
\lab{TMMTlambda}
\ee
which is still invariant under the global scale transformations \rf{scale-transf1}. To see application of this type of term in questions also related to DE, applied to a 
Gravity-Assisted Emergent Higgs Mechanism in the Post-Inflationary Epoch in \cite{Gravityassisted}.

In this context now the Weyl-rescaled metric ${\bar g}_{\m\n}$ that defines the Einstein frame  is given by
${\bar g}_{\m\n} = \chi_1  g_{\m\n}$ 
where
$\chi_1 \equiv \frac{\P_1 (A)}{\sqrt{-g}}$
one can indeed suspect that the extra term just induces a cosmological term in the Einstein frame, since using the previous  definition  we can check that  $ -2 \Lambda_0 \frac{\P_1 (A)^2}{\sqrt{-g}} = 
-2 \Lambda_0\sqrt{-{\bar g}}$.

This expectation  is indeed confirmed by a more detailed analysis, and we can see that the effective Lagrangian \rf{L-eff-final} is still valid, but now the corresponding $ A$ and $B$ coefficients are evaluated for  $\epsilon= 0$ and the effective potential is shifted by $2 \Lambda_0 $.

This has some consequences, like it is easier to guarantee a positive DE, just choose a big enough value for  $2 \Lambda_0 $. Apart from that there is not much change.

Notice also that $\P_1 (A)$ is  now not just a measure of integration, since it appears also square, and the theory with $\P_1 (A)$ appearing just linearly
has the highly non conventional infinite dimensional additional symmetry, valid for any regular set of four  functions $f^\delta $,  such that, 
$
A_{\mu\nu\gamma} \rightarrow A_{\mu\nu\gamma} + \epsilon_{\mu\nu\gamma\delta} f^\delta (R + L^{(1)})
$
(where $ \epsilon_{\mu\nu\gamma\delta}$  is the totally antisymmetric symbol taking values zero, one or minus one), which is
absent in the theory where $\P_1 (A)$ appears also squared, i.e. for $\Lambda_0 \neq 0$, so that can be an argument for  $\Lambda_0 = 0$, or a symmetry that protects 
$\Lambda_0 $ from becoming big after quantum corrections.

A stiff  era, as the one obtained in this work can have interesting observational consequences. This era could occurs before the BBN or after.  In both cases the stiff era can affect the light element abundances and perhaps the CMB, if the modes correspond to linear modes. Such an issue has been  discussed for example in \cite{2309.04850} and references therein. The study of this issue concerning the consequences of component with a stiff equation of state and how would be the special features that would be a consequence this stiff state in our scenario would be a very interesting subject for future research concerning our model.




\begin{acknowledgements}
E.G. want to thank the Pontificia  Universidad Cat\'olica de Valpara\'{\i}so, Chile,  for hospitality during this collaboration,  FQXi  for great financial support for work on this project at BASIC in Ocean Heights, Stella Maris, Long Island,  Bahamas  and  CA16104 - Gravitational waves, black holes and fundamental physics and  CA18108 - Quantum gravity phenomenology in the multi-messenger approach
for additional financial support.
\end{acknowledgements}


\begin{thebibliography}{10}
\bibitem{Guth}
Alan H. Guth, 
Phys.Rev.D {\bf23} (1981) 347-356; Adv.Ser.Astrophys.Cosmol. 3 (1987) 139-148 (reprint).

\bib{starobinsky}
A. Starobinsky, Phys.Lett. B {\bf91}, 99-102 (1980).

\bib{early-univ}
E.W. Kolb and M.S. Turner,  ``The Early Universe'', Addison Wesley (1990); \\
A. Linde,  ``Particle Physics and Inflationary Cosmology'', Harwood, Chur,
Switzerland (1990); \\
A. Guth,  ``The Inflationary Universe'', Vintage, Random House (1998); \\
S. Dodelson,  ``Modern Cosmology'', Acad. Press (2003);\\
S. Weinberg,  ``Cosmology'', Oxford Univ. Press (2008).
\bib{primordial}
V. Mukhanov,  ``Physical Foundations of Cosmology'', Cambridge Univ. Press (2005). 

\bib{emerg}
G.F.R. Ellis and R. Maartens, \CQG{21}{2004}{223} ~(\textsl{gr-qc/0211082}).

\bib{emerg2}
G.F.R. Ellis, J. Murugan and C.G. Tsagas,\CQG{21}{2004}{233} 
~(\textsl{ arxiv:gr-qc/0307112});\\
D.J. Mulryne, R. Tavakol, J.E. Lidsey and G.F.R. Ellis, 
\PRD{71}{2005}{123512} ~(\textsl{arxiv:astro-ph/0502589}); \\
A. Banerjee, T. Bandyopadhyay and S. Chaakraborty, {\sl Grav. Cosmol.} {\bf 13} 
(2007) 290-292 ~(\textsl{arxiv:0705.3933} [gr-qc]); \\
J.E. Lidsey and D.J. Mulryne \PRD{73}{2006}{083508}
~(\textsl{arxiv:hep-th/0601203}); \\
S. Mukherjee, B.C.Paul, S.D. Maharaj and A. Beesham, \textsl{arxiv:qr-qc/0505103}; \\
S. Mukherjee, B.C.Paul,  N.K. Dadhich, S.D. Maharaj and A. Beesham, 
\CQG{23}{2006}{6927} ~(\textsl{arxiv:gr-qc/0605134}).








\bib{accel-exp} 
M.S. Turner, in \textsl{Third Stromle Symposium ``The Galactic Halo''}, 
ASP Conference Series Vol.{\bf 666}, B.K. Gibson, T.S. Axelrod and M.E. Putman
(eds.), (1999); \\
N. Bahcall, J.P. Ostriker, S.J. Perlmutter and P.J. Steinhardt, \textsl{Science}  
{\bf 284},   (1999) 1481;\\
for a review, see P.J.E. Peebles and B. Ratra, {\sl Rev. Mod. Phys.} {\bf 75}, 
(2003) 559.
\bib{accel-exp-2}
A. Riess, {\em et al.}, \textsl{Astronomical Journal} {\bf 116} (1998) 1009-1038; \\ 
S. Perlmutter {\em et al.}, \textsl{Astrophysical Journal} {\bf 517} (1999) 565-586.
\bib{lambdaCDM}
A New cosmological paradigm: The Cosmological constant and dark matter
Lawrence M. Krauss, AIP Conf.Proc. {\bf444} (1998) 1, 59-69 • Contribution to: SILAFAE 98, 59-69, 5th International WEIN Symposium: A Conference on Physics Beyond the Standard Model (WEIN 98), Tropical Workshop on Particle Physics and Cosmology, PASCOS 1998 • e-Print: hep-ph/9807376 [hep-ph];
Ren-Yue Cen, Jeremiah P. Ostriker,  Astrophys.J. {\bf429} (1994) 4 • e-Print: astro-ph/9404012 [astro-ph].

\bib{H0}
 Jose Luis Bernal, Licia Verde, Adam G. Riess,	JCAP {\bf 10} (2016)019
DOI:	10.1088/1475-7516/2016/10/019, 	arXiv:1607.05617 [astro-ph.CO];
José Luis Bernal, Licia Verde, Raul Jimenez, Marc Kamionkowski, David Valcin et al.,  Phys.Rev.D {\bf103} (2021) 10, 103533 • e-Print: 2102.05066 [astro-ph.CO];
Leila L. Graef, Micol Benetti, Jailson S. Alcaniz,  Phys.Rev.D {\bf 99} (2019) 4, 043519 • e-Print: 1809.04501 [astro-ph.CO].

\bib{sigma8}
R.~E.~Keeley, S.~Joudaki, M.~Kaplinghat and D.~Kirkby,
JCAP \textbf{12} (2019), 035; 
K.~L.~Pandey, T.~Karwal and S.~Das,
JCAP \textbf{07} (2020), 026;
A.~Quelle and A.~L.~Maroto,
Eur. Phys. J. C \textbf{80} (2020) no.5, 369; A. Bhattacharyya, U. Alam, K. L. Pandey, S. Das, and S. Pal, Astrophys. J. 876, 143 (2019), arXiv:1805.04716 [astro-ph.CO];
G. Lambiase, S. Mohanty, A. Narang, and P. Parashari, Eur. Phys. J. C {\bf79}, 141 (2019), arXiv:1804.07154 [astroph.CO];
W.~Lin, K.~J.~Mack and L.~Hou,
Astrophys. J. Lett. \textbf{904} (2020) no.2, L22, arXiv:1910.02978 [astro-ph.CO]; M.~Berbig, S.~Jana and A.~Trautner,
Phys. Rev. D \textbf{102} (2020) no.11, 115008;
Quantifying the Sigma 8 tension with the Redshift Space Distortion data set,  David Benisty,  Phys.Dark Univ. {\bf31} (2021) 100766 • e-Print: 2005.03751 [astro-ph.CO].

\bib{slow-roll}
A. Linde, \PLB{108}{1982}{389-393}; \\
A. Albrecht and P. Steinhardt, \PRL{48}{1982}{1220-1223}.
\bib{slow-roll-param}
A.R. Liddle and D.H. Lyth, \PLB{291}{1992}{391-398} ~(\textsl{arxiv:astro-ph/9208007}); \\
A.R. Liddle and D.H. Lyth, \PR{231}{1993}{1-105} ~(\textsl{arxiv:astro-ph/9303019}).
\bib{peebles-vilenkin} 
P.J.E. Peebles and A.Vilenkin, \PRD{59}{1999}{063505}.

\bib{saitou-nojiri}
R. Saitou and S. Nojiri, \textsl{Eur. Phys. J.} {\bf C71} (2011) 1712
~(\textsl{arxiv:1104.0558} [hep-th]).
\bib{wetterich}
C. Wetterich, \PRD{89}{2014}{024005} ~(\textsl{arxiv:1308.1019} [astro-ph]).
\bib{murzakulov-etal}
Md. Wali Hossain, R. Myrzakulov, M. Sami and E.N. Saridakis, \PRD{90}{2014}{023512}
~(\textsl{arxiv:1402.6661} [gr-qc]).
 \bib{alphaatractors}
Konstantinos Dimopoulos, Charlotte Owen,  JCAP {\bf06} (2017) 027 • e-Print: 1703.00305 [gr-qc]; 
Konstantinos Dimopoulos, Leonora Donaldson Wood, Charlotte,  Phys.Rev.D {\bf97} (2018) 6, 063525 • e-Print: 1712.01760 [astro-ph.CO];R.~Herrera,
Eur. Phys. J. C \textbf{78}, no.3, 245 (2018); R.~Herrera,
Phys. Rev. D \textbf{98}, no.2, 023542 (2018);R.~Herrera,
Phys. Rev. D \textbf{99}, no.10, 103510 (2019); R.~Herrera,
Phys. Rev. D \textbf{102}, no.12, 123508 (2020); 
Llibert Aresté Saló, David Benisty, Eduardo I. Guendelman, Jaime d. Haro,  JCAP {\bf07} (2021) 007 • e-Print: 2102.09514 [astro-ph.CO];M.~Gonzalez-Espinoza, R.~Herrera, G.~Otalora and J.~Saavedra,
Eur. Phys. J. C \textbf{81}, no.8, 731 (2021);R.~Herrera and C.~Rios,
[arXiv:2210.10080 [gr-qc]];
Llibert Aresté Saló, David Benisty, Eduardo I. Guendelman,  Jaime d. Haro, Phys.Rev.D {\bf103} (2021) 12, 123535 • e-Print: 2103.07892 [astro-ph.CO].

\bib{Lorentzian}
David Benisty, Eduardo I. Guendelman,  Eur.Phys.J.C {\bf80} (2020) 6, 577 • e-Print: 2006.04129 [astro-ph.CO];
Lorentzian Quintessential Inflation, awarded 2nd prize in the 2020 Gravity Research Foundation Essays Competition, 
David Benisty, Eduardo I. Guendelman,  Int.J.Mod.Phys.D {\bf29} (2020) 14, 2042002 • e-Print: 2004.00339 [astro-ph.CO].



\bib{starobinsky-2}
S. Nojiri and S. Odintsov, \PRD{68}{2003}{123512} ~(\textsl{arxiv:hep-th/0307288}); \\
G. Cognola, E. Elizalde, S. Nojiri, S.D. Odintsov, L. Sebastiani and S. Zerbini,
\PRD{77}{2008}{046009} ~(\textsl{0712.4017} [hep-th]), and references therein;\\
S.A. Appleby, R.A. Battye and A.A. Starobinsky, 
\textsl{JCAP} {\bf 1006} (2010) 005 ~(\textsl{arxiv:0909.1737} [astro-ph]).


\bib{Re1}T.~P.~Sotiriou and V.~Faraoni,
Rev. Mod. Phys. \textbf{82}, 451-497 (2010)
doi:10.1103/RevModPhys.82.451
[arXiv:0805.1726 [gr-qc]]; S.~Nojiri and S.~D.~Odintsov,
Phys. Rept. \textbf{505}, 59-144 (2011)
doi:10.1016/j.physrep.2011.04.001
[arXiv:1011.0544 [gr-qc]]; S.~Capozziello and M.~De Laurentis,
Phys. Rept. \textbf{509}, 167-321 (2011)
doi:10.1016/j.physrep.2011.09.003
[arXiv:1108.6266 [gr-qc]]; R.~Herrera and N.~Videla,
Int. J. Mod. Phys. D \textbf{23}, no.08, 1450071 (2014)
doi:10.1142/S0218271814500710
[arXiv:1406.6305 [gr-qc]].

\bib{Re2} T.~Clifton, P.~G.~Ferreira, A.~Padilla and C.~Skordis,
Phys. Rept. \textbf{513}, 1-189 (2012)
doi:10.1016/j.physrep.2012.01.001
[arXiv:1106.2476 [astro-ph.CO]]; S.~Nojiri, S.~D.~Odintsov and V.~K.~Oikonomou,
Phys. Rept. \textbf{692}, 1-104 (2017)
doi:10.1016/j.physrep.2017.06.001
[arXiv:1705.11098 [gr-qc]]; S.~D.~Odintsov, V.~K.~Oikonomou, I.~Giannakoudi, F.~P.~Fronimos and E.~C.~Lymperiadou,
Symmetry \textbf{15}, no.9, 1701 (2023)
doi:10.3390/sym15091701
[arXiv:2307.16308 [gr-qc]].




\bib{TMT-orig-1}
E.I. Guendelman, \MPLA{14}{1999}{1043-1052} ~(\textsl{arxiv:gr-qc/9901017});\\
E.I. Guendelman, in  ``Energy Densities in the Universe'', Proc.
Rencontres de Moriond, Les Arcs (2000) ~(\textsl{arxiv:gr-qc/0004011}).
\bib{TMT-orig-2}
E.I. Guendelman and A. Kaganovich, \PRD{60}{1999}{065004}
~(\textsl{arxiv:gr-qc/9905029}).
\bib{TMT-orig-3}
E.I. Guendelman and O. Katz, \CQG{20}{2003}{1715-1728}
~(\textsl{arxiv:gr-qc/0211095}).
\bib{TMT-recent-1-a}
S. del Campo. E. Guendelman, R. Herrera and P. Labrana, \textsl{JCAP} {\bf 1006}
(2010) 026 ~(\textsl{arxiv:1006.5734} [astro-ph.CO]).
\bib{TMT-recent-1-b}
S. del Campo. E. Guendelman, A. Kaganovich, R. Herrera and P. Labrana, 
\PLB{699}{2011}{211} ~(\textsl{arxiv:1105.0651} [astro-ph.CO]).





\bib{TMT-recent-1-c}
E.I. Guendelman and P. Labrana, \IJMPD{22}{2013}{1330018}
~(\textsl{arxiv:1303.7267} [astro-ph.CO]).
\bib{TMT-recent-2}
E.I. Guendelman, D. Singleton and N. Yongram, \textsl{JCAP} {\bf 1211} (2012) 044
~(\textsl{arxiv:1205.1056} [gr-qc]);\\
E.I. Guendelman, H. Nishino and S. Rajpoot, \PLB{732}{2014}{156}
~(\textsl{arxiv:1403.4199} [hep-th]).

\bib{mstring}
 E.I. Guendelman, Class.Quant.Grav. {\bf17} (2000) 3673-3680 • e-Print: hep-th/0005041 [hep-th];
E.I. Guendelman, Phys.Rev.D {\bf63} (2001) 046006 • e-Print: hep-th/0006079 [hep-th]; 
E. Guendelman, A. Kaganovich, E. Nissimov and S. Pacheva, \PRD{66}{2002}{046003} 
~(\textsl{arxiv:hep-th/0203024}).
\bib{nishino-rajpoot}
H. Nishino and S. Rajpoot, \PLB{736}{2014}{350-355}
~(\textsl{arxiv:1411.3805} [hep-th]).

\bib{mstringspectrum}
International Journal of Modern Physics,  Implications of the spectrum of dynamically generated string tension theories E. I. Guendelman https://doi.org/10.1142/S0218271821420281, e-Print: 2110.09199 [hep-th].
\bib{mstringbranes}
Eduardo Guendelman,
Eur.Phys.J.C {\bf81} (2021) 10, 886 • e-Print: 2107.08005 [hep-th]; 
Eduardo Guendelman,
Eur.Phys.J.C {\bf82} (2022) 10, 857, DOI: 10.1140/epjc/s10052-022-10837-5.


\bib{susy-break}
E. Guendelman, E.Nissimov, S. Pacheva and M. Vasihoun, \textsl{Bulg. J. Phys.}
{\bf 40} (2013) 121-126 ~(\textsl{arxiv:1310.2772} [hep-th]); \\
E. Guendelman, E.Nissimov, S. Pacheva and M. Vasihoun, \textsl{Bulg. J. Phys.}
{\bf 41} (2014) 123-129 ~(\textsl{arxiv:1404.4733} [hep-th]).

\bib{ourquintessence} 
Eduardo Guendelman, Ramón Herrera, Pedro Labrana, Emil Nissimov, Svetlana Pacheva,  Gen.Rel.Grav. {\bf47} (2015) 2, 10 • e-Print: 1408.5344 [gr-qc].


\bib{quintess}
E. Guendelman, E. Nissimov and S. Pacheva, \textsl{arxiv:1407.6281} [hep-th].




\bib{curv}Eduardo I. Guendelman, Ramon Herrera,  Gen.Rel.Grav. {\bf48} (2016) 1, 3 • e-Print: 1511.08645 [gr-qc].
\bibitem{curv2}  Eduardo I. Guendelman, Ramon Herrera, Pedro Labrana,  Phys.Rev.D {\bf 103} (2021) 123515 • e-Print: 2005.14151 [gr-qc] and references there.


\bib{ourquintessencewithEDE}
Eduardo Guendelman, Ramón Herrera, David Benisty, Phys.Rev.D {\bf105} (2022) 12, 124035 • e-Print: 2201.06470 [gr-qc].













%
 
 
 
 \bibitem{Gravityassisted}
Eduardo Guendelman, Emil Nissimov, Svetlana Pacheva, Int.J.Mod.Phys.D {\bf25} (2016) 12, 1644008 • e-Print: 1603.06231 [hep-th].
 


\bibitem{Kaiser:2013sna}
D.~I.~Kaiser and E.~I.~Sfakianakis,
Phys. Rev. Lett. \textbf{112}, no.1, 011302 (2014)
doi:10.1103/PhysRevLett. 112.011302
[arXiv:1304.0363 [astro-ph.CO]].








\bibitem{p1} X.~Chen, M.~X.~Huang, S.~Kachru and G.~Shiu,
\textsl{JCAP} {\bf 0701} (2007) 002; B.~A.~Bassett, S.~Tsujikawa and D.~Wands,
Rev. Mod. Phys. \textbf{78},  (2006) 537-589.


\bibitem{BICEP:2021xfz}
P.~A.~R.~Ade \textit{et al.} [BICEP and Keck],
Phys. Rev. Lett. \textbf{127} (2021) no.15, 151301
doi:10.1103/PhysRevLett.127.151301
[arXiv:2110.00483 [astro-ph.CO]].

\bibitem{Planck:2018vyg}
N.~Aghanim \textit{et al.} [Planck],
Astron. Astrophys. \textbf{641}, A6 (2020)
[erratum: Astron. Astrophys. \textbf{652}, C4 (2021)]
doi:10.1051/0004-6361/201833910
[arXiv:1807.06209 [astro-ph.CO]].

\bibitem{Scherrer:2004au}
R.~J.~Scherrer,
Phys. Rev. Lett. \textbf{93}, 011301 (2004)
doi:10.1103/PhysRevLett.93.011301
[arXiv:astro-ph/0402316 [astro-ph]].

\bibitem{guendelmanKineticDEDM}
Eduardo Guendelman, Emil Nissimov, Svetlana Pacheva, 
Eur.Phys.J.C {\bf76} (2016) 2, 90 • e-Print: 1511.07071 [gr-qc].

\bibitem{Frieman:2008sn}
J.~Frieman, M.~Turner and D.~Huterer,
Ann. Rev. Astron. Astrophys. \textbf{46}, 385-432 (2008).


\bibitem{Libro}
 M. Abramowitz, I.A. Stegun (eds.), {\it{Handbook of Mathematical
Functions with Formulas, Graphs, and Mathematical Tables}}, 9th
printing (Dover, New York, 1972).














 
 



















\bibitem{Ar1}
I. Zlatev, L.-M. Wang, and P. J. Steinhardt, Phys. Rev.
Lett. {\bf82}, 896 (1999).
\bibitem{Ar2} P. J. Steinhardt, L.-M. Wang, and I. Zlatev, Phys. Rev.
D {\bf59}, 123504 (1999);
 I. Zlatev and P. J. Steinhardt, Phys. Lett. B {\bf459}, 570
(1999).

\bibitem{Chavanis:2014lra}
P.~H.~Chavanis,
Phys. Rev. D \textbf{92}, no.10, 103004 (2015).


\bibitem{Nojiri:2006jy}
S.~Nojiri, S.~D.~Odintsov and H.~Stefancic,
Phys. Rev. D \textbf{74}, 086009 (2006)
doi:10.1103/PhysRevD.74.086009
[arXiv:hep-th/0608168 [hep-th]].


\bibitem{sloth}
Florian Niedermann, Martin S. Sloth. 
Phys.Lett.B {\bf835} (2022) 137555 • e-Print: 2112.00759 [hep-ph].
\bibitem{sloth2}
Juan S. Cruz, Florian Niedermann, Martin S. Sloth
• e-Print: 2209.02708 [astro-ph.CO].

\textcolor{red}{\bibitem{2309.04850}
A Stiff Pre-CMB Era with a Mildly Blue-tilted Tensor Inflationary Era can Explain , the 2023 NANOGrav Signal, V.K. Oikonomou, arXiv:2309.04850.}





\end{thebibliography}
\end{document}